\documentclass[10pt,a4paper]{article}
\usepackage[a4paper,left=2.5cm,right=2.5cm,top=3cm,bottom=3cm]{geometry}
\usepackage[utf8]{inputenc}

\usepackage{cite}

\usepackage{nameref,hyperref}

\usepackage{microtype}
\DisableLigatures[f]{encoding = *, family = * }

\usepackage{changepage}

\usepackage[aboveskip=1pt,labelfont=bf,labelsep=period,singlelinecheck=off]{caption}

\makeatletter
\renewcommand{\@biblabel}[1]{\quad#1.}
\makeatother

\usepackage{lastpage,fancyhdr,graphicx}
\usepackage{epstopdf}
\fancyheadoffset[L]{-0in}
\fancyfootoffset[L]{-0in}

\usepackage{color}

\definecolor{Gray}{gray}{.25}

\usepackage{graphicx}

\usepackage{sidecap}

\usepackage{wrapfig}
\usepackage[pscoord]{eso-pic}
\usepackage[fulladjust]{marginnote}

\usepackage{amsmath}
\usepackage{subfig}
\usepackage{multirow}
\usepackage{float}
\usepackage{indentfirst}
\usepackage[normalem]{ulem}
\usepackage{fancyhdr}
\fancypagestyle{ttlpage}{
	\fancyhf{}
	\fancyhead[L]{ \bfseries \textcolor{red}{\Large This is a preprint of an open-acess article published in Journal of Geodesy. The final authenticated version is available online at: \url{https://doi.org/10.1007/s00190-021-01498-5}}}
}

\reversemarginpar

\begin{document}
	\vspace*{0.35in}
	\thispagestyle{ttlpage}
	\begin{flushleft}
		{\Large
			\textbf\newline{Revisiting the Light Time Correction in Gravimetric Missions Like GRACE and GRACE Follow-On}
		}
		\newline
		\\
		Yihao Yan\textsuperscript{1,2,**},
		Vitali M\"uller\textsuperscript{3,*},
		Gerhard Heinzel\textsuperscript{3},
		Min Zhong\textsuperscript{4},
		\\
		\bigskip
		\bf{1} School of Physics, Huazhong University of Science and Technology, Wuhan, 430074, China
		\\
		\bf{2} Institute of Geodesy and Geophysics, Chinese Academy of Sciences, Wuhan, 430077, China
		\\
		\bf{3} Max-Planck-Institut f\"ur Gravitationsphysik (Albert-Einstein-Institut) and Institut f\"ur Gravitationsphysik of Leibniz Universit\"at Hannover, 30167 Hannover, Germany
		\\
		\bf{4} School of Geospatial Engineering and Science, Sun Yat-Sen University, Zhuhai, 519082, China
		\\
		\bigskip
		*  vitali.mueller@aei.mpg.de\\
		** yihaoyan@hust.edu.cn
		
	\end{flushleft}
\newcommand\myfig[1]{\includegraphics[width=0.48\textwidth]{#1}}
\newcommand\vdp{.}
\newcommand\reviewermark[1]{#1}  %
\newcommand\reviewersout[1]{}  %

\section*{Abstract}
The gravity field maps of the satellite gravimetry missions GRACE (Gravity Recovery and Climate Experiment) and GRACE Follow-On are derived by means of precise orbit determination. The key observation is the biased inter-satellite range, which is measured primarily by a K-Band Ranging system (KBR) in GRACE and GRACE Follow-On. The GRACE Follow-On satellites are additionally equipped with a Laser Ranging Interferometer (LRI), which provides measurements with lower noise compared to the KBR. The biased range of KBR and LRI needs to be converted for gravity field recovery into an instantaneous range, i.e.~the biased Euclidean distance between the satellites' center-of-mass at the same time. One contributor to the difference between measured and instantaneous range arises due to the non-zero travel time of electro-magnetic waves between the spacecraft. 
We revisit the calculation of the light time correction (LTC) from first principles considering general relativistic effects and state-of-the-art models of Earth's potential field. The novel analytical expressions for the LTC of KBR and LRI can circumvent numerical limitations of the classical approach. The dependency of the LTC on geopotential models and on the parameterization is studied, and afterwards the results are compared against the LTC provided in the official datasets of GRACE and GRACE Follow-On. It is shown that the new approach has a significantly lower noise, well below the instrument noise of current instruments, especially relevant for the LRI, and even if used with kinematic orbit products. This allows calculating the LTC accurate enough even for the next generation of gravimetric missions.
\section*{Keywords}
GRACE Follow-On $\cdot$ light time correction $\cdot$ general relativity $\cdot$ Laser interferomery $\cdot$ K-Band Ranging


	\section{Introduction}
\label{intro}
The twin GRACE satellites observed Earth's gravity field and, more importantly, the monthly time variations of it from the launch in 2002 until their reentry in 2017. These variations reflect the mass transport on large scale in and on Earth. The measurement principle is based  on low-low satellite-satellite tracking (LL-SST), i.e.~measuring distance variations between the orbiters, which are separated on the same polar orbit by approx.~200\,km \cite{tapley2004gravity}. The inter-satellite range variations were measured by the K-Band Ranging system (KBR) with a noise level of approx.~1\,$\mu$m/$\sqrt{\textrm{Hz}}$ at a Fourier frequency of 0.1\,Hz, and with elevated noise towards lower frequencies.

Due to the enormous success of GRACE, a successor mission called GRACE Follow-On (GFO) was launched on May 22, 2018.   Its payload, an evolved version of the original GRACE,  is comprised of, among others, GNSS~receivers for precise orbit determination, accelerometers for the measurement of non-gravitational accelerations, \reviewermark{star cameras and inertial measurement units for attitude determination and the aforementioned KBR system} \cite{kornfeld2019grace}. In addition, GRACE Follow-On hosts the novel Laser Ranging Interformeter (LRI) which is a technology demonstrator, and it is the first inter-satellite laser interferometer in space. It has demonstrated an excellent performance and reliability of all subsystems and exhibits a noise level of approx.~1\,nanometer/$\sqrt{\textrm{Hz}}$ at a Fourier frequency of 0.1\,Hz, well below the requirements \cite{abich2019orbit}. The novel LRI and conventional KBR are operated in parallel and, \reviewermark{since both should measure the same Euclidean distance variations after some post-processing corrections that are described below}, inter-comparisons and cross-calibrations can be performed in order to characterize the instruments and their behavior.

Both instruments rely on the transmission of electro-magnetic radiation, back and forth, between the satellites. The LRI operates at an optical frequency of $\approx$\,281\,THz in a so-called active transponder configuration \cite{sheard2012intersatellite}, while the KBR, often also called the microwave ranging instrument (MWI), uses two microwave frequencies, one in the K and one in the Ka band, in the so-called dual one-way ranging (DOWR) scheme \cite{kim2000simulation, thomas1999analysis}. Both instruments rely on tracking the phase of a beatnote signal at low radio frequencies ($\leq 18$\,MHz). The tracked phase is - up to an unknown offset - proportional to the travel time of the radiation between the orbiters, thus, proportional to the inter-satellite distance variations from an initial epoch where phase tracking started. When the phase measurements are rescaled to a displacement, they are usually referred to as biased range observations \reviewermark{in the official data products}. 

The gravity field recovery algorithms usually are based on the corrected \reviewermark{(i.e.~ instantaneous Euclidean)} biased range \reviewermark{or on its time derivative called range rate \cite{naeimi2017global}. The former one means} the Euclidean biased distance between both satellites' center-of-mass at the same epoch, which differs from the measured biased range due to effects from the finite speed of light and due to the fact that the measurements are not referred to the center-of-mass. The difference between biased and \reviewermark{corrected} instantaneous range is usually expressed as the sum of three terms: the light-time correction, the ionospheric correction, and the antenna phase center correction - often called tilt-to-length coupling in the context of laser interferometry. 

The LRI was designed to have a minimal tilt-to-length coupling, which has been confirmed by in-flight measurements to be below 150 $\mu$m/rad \cite{wegener2020}. The coupling is significantly lower than for the KBR \cite{wu2006algorithm}, where the reference point for the range measurement is offset by approx.~1.4\,m from the center-of-mass. The ionospheric effect is also insignificant in the case of the LRI due to the shorter wavelength of the optical radiation. \reviewermark{The ionospheric correction for the KBR is briefly addressed in this paper, mainly to show that there is a cross-coupling of ionospheric effect and light time correction (LTC), but it is highly suppressed to a level below picometers in the employed two-way measurements. The main focus of this paper} lies on the LTC, which is relevant for KBR and LRI and which was mentioned first for the GRACE satellites in \cite{thomas1999analysis}. Later, \cite{kim2000simulation} described a method  to analytically calculate the light time correction based on absolute spacecraft velocities, i.e.~only the special relativistic contribution. \cite{turyshev2014general} established an extensive description of general relativistic observables in GRACE-like missions, which includes an analytical model for the light-time correction, among others. However, in our opinion, it is not straightforward to apply the formalism to actual flight data, \reviewermark{ because the relevant LTC expressions are derived under the assumption of nearly-matched Keplerian orbits for the satellites and approximations are used to derive closed-form expressions for the LTC. This enables the authors to understand and discuss the individual terms, but is also a restriction with regard to generality. }

Thus, we derive the light time correction from first principles, and stay close to the data products and processing strategy in gravimetric mission, such that the results are easily applicable. The potential of Earth's gravity field is expressed in terms of Stokes coefficients of a spherical harmonic (SH) expansion and the equations are formulated with quantities available from the official public data of the missions. In the following sec.~\ref{sec:eom}, the equations of motion are introduced in the general relativistic context, which are needed to describe the propagation of electro-magnetic waves. The propagation time of light between satellites is derived and split into the contributions from relativity (sec.~\ref{sec:ltc_rel}) and atmosphere (sec.~\ref{sectmedia}). However, actual calculations require a solution of an implicit equation (sec.~\ref{secLightTimeEquation}), which can be solved iteratively or by means of an analytical approximation. The analytical approach offers some advantages, since it allows us to replace some orbit product quantities that drive the numerical precision with more precise ranging observations. The analytical solution is combined in sec.~\ref{secLTCinDOWR} into the dual-way light time corrections for KBR and in sec.~\ref{secLTCinTWR} into the round-trip LTC for LRI. Sec.~\ref{secRequirements} addresses the sensitivity of the ranging instruments and sketches a potential goal for the precision of the analytical equations and background models for the LTC. 
In the subsequent section~\ref{secValidationOWR}, the analytical expressions for the one-way LTC are verified against numerical results and a parameter study is performed regarding background model accuracy and degree of approximations.
We compare our results for the LTC against the results from official datasets for GRACE and GRACE Follow-On in sec.~\ref{sec:Comparison}, while sec.~\ref{sec:Enhancing} addresses further potential improvements in the light-time correction calculation.

\section{Equations of Motion in General Relativity}
\label{sec:eom}
\indent In order to derive a precise light time correction, the travel time of light between satellites is needed in a general relativistic context. For this, it is convenient to describe the light or microwaves in terms of mass-less particles, the photons, which move on geodesics according to the equations of motion in general relativity. We denote the coordinates of an object in the Geocentric Celestial Reference System (GCRS) as:
\begin{equation} 
x^{\alpha}=\left(c_{0} \cdot t, x, y, z\right)=\left(c_{0} \cdot t, \vec{r}\right)=\left(x^{0}, x^{1}, x^{2}, x^{3}\right)^{\top}
\end{equation}  
where the common four-vector notation from relativity is used, and $c_{0}$ is the proper speed of light for vacuum in a local Lorentz frame with a numerical value of 299 792 458 m/s, $\vec{r}$ is the three dimensional spatial vector.  \\
\indent We employ the sign convention $\gamma_{\alpha \beta} = \textrm{diag} \{-1, +1, +1, +1\}$ for the Minkowski metric as used, for instance, by \cite{kopeikin2011relativistic}. The Greek indices such as $\alpha$ and $\beta$ range from 0..3, while Latin letters like \textit{m} and \textit{n} denote spatial components and range from 1..3. $c_{n}$ is the coordinate speed of light.

The metric tensor $g_{\alpha \beta}$ of the Earth in the GCRS is approximated by a Post-Newtonian expansion as \cite{turyshev2014general,soffel2012space}:
\begin{equation}
\begin{aligned} g_{00} &=\gamma_{00} + \frac{2 W}{c_{0}^{2}}-\frac{2 W^{2}}{c_{0}^{4}}+\mathcal{O}\left(c_{0}^{-6}\right) \\ g_{0m} &=g_{m0}=-\frac{4 \vec{V}_m}{c_{0}^{3}}+\mathcal{O}\left(c_{0}^{-5}\right) \\ g_{m m} &=\gamma_{mm}+\frac{2 W}{c_{0}^{2}}+\mathcal{O}\left(c_{0}^{-4}\right) \end{aligned}
\label{eqMetricTensor}
\end{equation}
with
\begin{equation}
W = W_{e} + \sum_i W_{\textrm{cb},i}
\end{equation}
where $W_e$ is the classical Newtonian potential due to the mass distribution of the Earth. Moreover, $W$ contains a sum of potentials $W_{\textrm{cb},i}$ giving rise to the direct tidal acceleration towards other celestial bodies, in particular the Sun and the Moon. The vector potential $\vec{V} $ in eq.~(\ref{eqMetricTensor}) accounts for Earth's spin moment with $\vec{V}_m $ denoting the \textit{m}th component of $\vec{V} $.

We describe the potential $W_{e}$ as the sum of a central term $W_\textrm{PM} = GM_e/r$ and of higher moments of the gravity field $W_\textrm{HM}$, i.e.~
\begin{equation}
W_{e} = W_\textrm{PM} + W_\textrm{HM} = W_\textrm{PM} + W_\textrm{G} + W_\textrm{tidal} +W_\textrm{non-tidal},
\end{equation}
whereby $W_\textrm{HM}$ is formed by the higher moment of static mass distribution potential $W_G$, by the \reviewersout{tidal} potential $W_\textrm{tidal}$ describing  the distortion of the mass distribution due celestial bodies such as Moon and Sun, and by the non-tidal potential $W_\textrm{non-tidal}$ describing small variations in the atmosphere, oceans, hydrology, ice and solid earth (AOHIS). These non-tidal variations contain highly interesting information for Earth sciences and the measurement of them is the main objective of GRACE-like missions.

The potentials describing higher moments of the gravity field are usually expressed in terms of a SH expansion \cite{heiskanenh}:
\begin{equation}
W_\textrm{HM}(r, \Theta,\lambda)=\frac{G M_{e}}{R_{e}} \sum_{l=1}^{\infty}\left(\frac{R_{e}}{r}\right)^{(l+1)} \sum_{k=0}^{l}\left(\overline{C}_{l k} \cos (k \lambda) +\overline{S}_{l k} \sin (k \lambda) \right) \overline{P}_{l k}(\cos \Theta)
\end{equation}
where G is the gravitational constant, $M_{e}$ is the mass of the Earth, $R_{e}$ is Earth's average radius,
$(r, \Theta,\lambda)$ are the spherical position coordinates, $\overline{P}_{l k}$ are the normalized Legendre functions of the second kind, \textit{l} and \textit{k} are the degree and order of the series expansion, and $\overline{C}_{l k}$ and $\overline{S}_{l k}$ are the normalized dimensionless Stokes coefficients. The Stokes coefficients of the static, tidal and non-tidal models given in table~\ref{tab:1} can be summed up in order to yield the total field $W_\textrm{HM}$.  

\begin{table}
	\centering	
	\caption{List of background models used in calculations}
	\label{tab:1} 
	\begin{tabular}{lll}
		\hline\noalign{\smallskip}
		Potential & Abbreviation&Parameters or Model\\
		\noalign{\smallskip}\hline\noalign{\smallskip}
		Static gravity field & STG&GGM05s \cite{ries2016development} \\	
		Solid earth tides& SET& IERS 2010 \cite{petit2010iers} \\ 
		Ocean tides &OT&EOT11a \cite{savcenko2012eot11a} \\
		Pole tides&PT&IERS 2010 \cite{petit2010iers} \\
		Ocean pole tides&PT&Desai 2003 \cite{petit2010iers} \\
		Atmospheric tides (S1, S2)&AT &Bode-Biancale 2003 \cite{biancale2006mean} \\
		Atmosphere and Ocean Dealiasing&AOD & AOD1B RL06 \cite{dobslaw2017new} \\
		Celestial Body &SunMoon&DE421\cite{folkner2008planetary} \\
		\noalign{\smallskip}\hline
	\end{tabular}
\end{table}

The direct acceleration towards a celestial body, which is often called direct tidal acceleration, has in the Earth-centered frame the potential $W_{\textrm{cb},i}$ \cite{montenbruck2002satellite}:
\begin{equation}
W_{\textrm{cb},i}=\frac{G M_{\textrm{cb},i}}{R_{\textrm{cb},i}} \sum_{l=2}^{\infty}\left(\frac{r}{R_{\textrm{cb},i}}\right)^{l} \overline{P}_{l}(\cos \varsigma_i)
\end{equation}
where \textit{G} is the gravitational constant, $M_{\textrm{cb},i}$ is the mass the of $i$-th celestial body, $R_{\textrm{cb},i}$ is the distance between Earth and celestial body, $r$ is the distance between Earth center and the satellite, $\overline{P}_{l}$ are the normalized Legendre functions of the first kind, $\varsigma_i$ is the angle between $\vec{R}_{\textrm{cb},i}$ and the satellite position vector $\vec{r}_{s}$, and \textit{l} is the degree of the series expansion. In this paper we consider only the Sun and the Moon, since they are dominating the direct tidal acceleration.

\indent The vector potential $\vec{V} $ in eq.~(\ref{eqMetricTensor}) is usually approximated as \cite{turyshev2014general}:
\begin{equation}
\vec{V}(t, \vec{r}) \approx \frac{G M_{e}}{2 \cdot r^{3}} \cdot \vec{S} \times \vec{r}+\mathcal{O}\left(x^{-4}, c^{-2}\right)
\end{equation}
where $\vec{S}$ is Earth's spin moment, or its angular momentum per unit of mass. It can be approximated by the angular momentum of a homogeneous sphere:
\begin{equation}
\vec{S} \approx \frac{2}{5} \cdot R_{e}^{2} \cdot \vec{\omega}_{e}
\end{equation}
where $\vec{\omega}_{e}$ is Earth's angular velocity vector. \\

The equations of motions of a point particle, e.g. satellites or light read in the context of General Relativity as \cite{kopeikin2011relativistic}:
\begin{equation}
\frac{ \textrm{d}^2 x^k }{ \textrm{d} t^2 } = -\Gamma^{k}_{~\alpha \beta} \cdot \frac{\textrm{d} x^\alpha}{\textrm{d}  t} \cdot \frac{\textrm{d} x^\beta}{\textrm{d}  t} + \frac{1}{c_0} \Gamma^{0}_{~\alpha \beta} \cdot \frac{\textrm{d} x^\alpha}{\textrm{d}  t} \cdot \frac{\textrm{d} x^\beta}{\textrm{d}  t} \cdot \frac{\textrm{d} x^k}{\textrm{d}  t} \quad \textrm{with} \quad k = 1..3,
\label{eqGRequationmotion}
\end{equation}
where $t$ is the coordinate time, and $\Gamma^{k}_{~\alpha \beta}$ are the Christoffel symbols, which depend on derivatives of the metric tensor $g_{\alpha \beta}$. It is straightforward to numerically integrate these differential equations in order to obtain a trajectory for a given set of initial conditions. For a photon, the trajectory appears bent with approximately twice the classical Newtonian acceleration towards Earth's center, consistent with one of the very early results of GR \cite{Einstein:1911vc, Soldner}. The selection of the initial velocity of a photon requires the coordinate speed of light, which depends on the metric tensor and on the propagation direction. It can be derived from the following ansatz for the four velocity:
\begin{equation}
\frac{\mathrm{d} x^{\alpha}}{\mathrm{d} t}=\left(c_{0}, \vec{d_{0}} \vdp c_{n}\right)^{\mathrm{T}}
\end{equation}
where $ c_{n}$ is the coordinate speed of light in a vacuum in the GCRS, $\vec{d_{0}}$ is the normalized propagation direction of the photon and $t$ is the coordinate time of the GCRS. 

\indent The interval $\textrm{d}s^2$ of a world line or trajectory of a massless particle vanishes \cite{kopeikin2011relativistic}: 
\begin{equation}
\mathrm{d} s^{2}=g_{\alpha \beta}(t, \vec{r}) \cdot \mathrm{\partial} x^{\alpha} \cdot \mathrm{\partial} x^{\beta}=0.
\label{eqds2}
\end{equation}
After dividing by $\textrm{d}t^{2}$ and plugging eq.~(\ref{eqMetricTensor}) into eq.~(\ref{eqds2}), one obtains a quadratic equation for $c_n$
\begin{equation}
\begin{aligned} 0 &=g_{\alpha \beta}(t, \vec{r}) \cdot \mathrm{d} x^{\alpha} / \mathrm{d} t \cdot \mathrm{d} x^{\beta} / \mathrm{d} t \\ &=c_{0}^{2} \cdot g_{00}+\vec{G} \vdp \vec{d_{0}} \cdot c_{n} \cdot c_{0}+c_{n}^{2} \cdot g_{mm}, \end{aligned}
\end{equation}
where $\vec{G}=2(g_{01},g_{02},g_{03})^{T}=-8\vec{V}/c_{0}^{3}$ and $g_{mm} = g_{11}= g_{22}= g_{33}$.  The post-Newtonian effect is very small,  such that  $g_{00}$ and $g_{mm}$ are close to unity. The quadratic equation can be solved and the solution with positive propagation velocity is taken for the coordinate speed of light:
\begin{equation}
c_{n}=c_{0} \cdot \sqrt{-\frac{g_{00}}{g_{mm}}+\frac{\left(\vec{G} \vdp \vec{d}_{0}\right)^{2}}{4 \cdot\left(g_{mm}\right)^{2}}}-c_{0} \cdot \frac{\vec{G} \vdp \vec{d}_{0}}{2 \cdot\left(g_{mm}\right)} =\sqrt{\frac{c_{0}^{6}-2 \cdot c_{0}^{2} \cdot W^{2}+4 \cdot W^{3}+16 \cdot\left(\vec{V} \vdp \vec{d}_{0}\right)^{2}}{\left(c_{0}^{2}+2 \cdot W\right)^{2}}}+\frac{4 \cdot \vec{V} \vdp \vec{d}_{0}}{c_{0}^{2}+2 \cdot W}. \label{eqCnFinal}
\end{equation}

\indent The infinitesimal propagation time $\textrm{d}t$ of a photon is related to the coordinate pathlength $\textrm{d}s$ through
\begin{equation}
\textrm{d}t = \frac{n}{c_{n} } \cdot \textrm{d}s = \frac{1+2 \cdot W / c_{0}^{2}-4 \cdot \vec{V} \vdp \vec{d}_{0} / c_{0}^{3}}{c_{0}} \cdot n \cdot \textrm{d}s+\mathcal{O}\left(c_{0}^{-5}\right),
\end{equation}
where $n$ denotes the refractive index at the location of the photon.

\indent For a one-way ranging measurement, the propagation time $\varDelta t $ of a photon traveling along path $\mathcal P$ can be written as
\begin{equation}
\varDelta t=\int_{\mathcal{P}} \frac{n}{c_{n}\left(t, \vec{r}_{\mathrm{ph}}\right)} \mathrm{d} s \approx \underbrace{\int_{\mathcal{P}} \frac{1}{c_{n}\left(t, \vec{r}_{\mathrm{ph}}\right)} \mathrm{d} s}_{\Delta t_\textrm{rel}} + \underbrace{\frac{1}{c_0} \int_{\mathcal{P}} (n-1)~\mathrm{d} s}_{\Delta t_\textrm{media}},
\label{eqDeltaT1}
\end{equation}
where $\vec{r}_{\mathrm{ph}}$  is the position of the photon on the path $\mathcal{P}$ and $t$ is the coordinate time. Since $c_n$ is close to $c_0$ and since the effect of the refractive index due to the ionospheric and neutral atmosphere is small, such that $(n-1)$ is close to zero, it is possible to approximate the integral as the sum of the relativistic effect ($\Delta t_\textrm{rel}$) and a contribution from the refractive index of the media ($\Delta t_\textrm{media}$). Both effects are analyzed in more detail in the next two subsections.

\section{Light time correction $\Delta t_\textrm{rel}$ due to relativity}
\label{sec:ltc_rel}
The light path $\mathcal{P}$ between satellites in a gravimetric mission can be assumed as a straight line in the three-dimensional coordinate system, which can be parameterized by a parameter $\lambda \in [0,1]$:
\begin{equation}
\vec{r}_{\mathrm{ph}}(\lambda)=\vec{r}_{e}+\left(\vec{r}_{r}-\vec{r}_{e}\right) \cdot \lambda
\end{equation}
where $\vec{r}_{r}$ is the three-dimensional position of the photon reception and $\vec{r}_{e} $ is the three-dimensional  position of the photon emission.

This neglects the relativistic light bending, which arises from an apparent acceleration $a_c$ of the photons towards the geocenter with twice the Newtonian acceleration \cite{kopeikin2011relativistic}, i.e.~$a_c = 2GM_{e}/r^2 $. 
The displacement of a photon in radial direction w.r.t.~a straight line is of the order of $a_c \cdot (\Delta t)^2 / 2 \approx 4~\mu m$, where a propagation time of $\Delta t = 200~\textrm{km} / c_0 \approx 0.66~\textrm{msec} $ and a satellite position of $r = 6731$ km was assumed. Temporal variations of the displacement due to higher moments of the gravity field are much smaller. In the domain of phasefronts, the light-bending yields a negligible static phasefront tilt of the order of $4~\mu \textrm{m}/200~\textrm{km} \approx 2 \cdot 10^{-11} \textrm{rad}$.

Thus, one can anticipate that the light-time correction derived from the bent light path will differ only insignificantly from a correction derived on the straight line. The approximation is further justified in sec.~\ref{secValidationOWR}, where our simplified analytical results are compared to results obtained via numerically integrating eq.~(\ref{eqGRequationmotion}) and thus, accounting for the full GR effects.

\indent Evaluating the propagation time $\varDelta t_\textrm{rel}$ in eq.~(\ref{eqDeltaT1}) with the photon path $\mathcal{P}$ yields:
\begin{align}
\varDelta t_\textrm{rel}&=\int_{\lambda=0}^{1} c_{n}^{-1}\left(t, \vec{r}_{\mathrm{ph}}\right) \cdot\left|\frac{\mathrm{d} \vec{r}_{\mathrm{ph}}}{\mathrm{d} \lambda}\right| \mathrm{d} \lambda =\left|\vec{r}_{e}-\vec{r}_{r}\right| \cdot \int_{\lambda=0}^{1} c_{n}^{-1}\left(t, \vec{r}_{\mathrm{ph}}\right) \mathrm{d} \lambda \\
&\approx \underbrace{\frac{\left|\vec{r}_{r}-\vec{r}_{e}\right|}{c_{0}}}_{\varDelta t_\textrm{SR}}
+ \underbrace{2 \cdot \varDelta t_\textrm{SR} \cdot \int_{\lambda=0}^{1} \frac{G M_{e}}{c_{0}^{2} \cdot\left|\vec{r}_{\mathrm{ph}}(\lambda)\right|} d \lambda}_{\mathcal{T}_{\mathrm{PM}}}
+\underbrace{2 \cdot \varDelta t_{\mathrm{SR}} \cdot \int_{\lambda=0}^{1} \frac{W_{\mathrm{HM}}\left(	t(\lambda), \vec{r}_{\mathrm{ph}}(\lambda)\right)}{c_{0}^{2}} \mathrm{d} \lambda}_{\mathcal{T}_{\mathrm{HM}}} \nonumber \\ &\quad + \underbrace{\varDelta t_{\mathrm{SR}} \cdot \int_{\lambda=0}^{1} \frac{-4 \cdot \vec{V}\left(\vec{r}_{\mathrm{ph}}(\lambda)\right) \vdp \vec{d}_{0}}{c_{0}^{3}} \mathrm{d} \lambda}_{\mathcal{T}_{ \mathrm{SM}}}, \label{eqDeltaT2}
\end{align}
where terms with the order of $c_{0}^{-4}$ and smaller were omitted and where 
the normalized propagation direction of the photon $\vec{d_{0}}$ was abbreviated by
\begin{equation}
\vec{d}_{0}=\frac{\vec{r}_{r}-\vec{r}_{e}}{\left|\vec{r}_{r}-\vec{r}_{e}\right|}.
\label{eqd0def}
\end{equation}

\indent  In upper eq.~(\ref{eqDeltaT2}), the first term $\Delta t_\mathrm{SR}$ is the propagation time from special relativity in flat space-time, the second term $\mathcal{T}_\mathrm{PM}$ is the time delay due to Earth's central field, the third term $\mathcal{T}_\mathrm{HM}$ is the time delay from higher moments of the gravitational potential due to Earth's mass distribution and due to other celestial bodies, and the fourth term $\mathcal{T}_\mathrm{SM}$ is  the time delay due to Earth's spin moment. \\
\indent The term $\mathcal{T}_{\mathrm{PM}}$ is commonly called Shapiro time delay and it has a closed analytical form \cite{turyshev2014general}
\begin{equation}
\mathcal{T}_{\mathrm{PM}}=\frac{2 \cdot G M_{e}}{c_{0}^{3}} \cdot \ln \left(\frac{\left|\vec{r}_{r}\right|+\vec{d_{0}} \vdp \vec{r}_{r}}{\left|\vec{r}_{e}\right|+\vec{d}_{0} \vdp \vec{r}_{e}}\right)
= \frac{2 \cdot G M_{e}}{c_{0}^{3}} \cdot \ln \left(\frac{\left|\vec{r}_{r}\right|+\left|\vec{r}_{e}\right|+\left|\vec{r}_{r}-\vec{r}_{e}\right|}{\left|\vec{r}_{r}\right|+\left|\vec{r}_{e}\right|-\left|\vec{r}_{r}-\vec{r}_{e}\right|}\right). \label{eqTPM}
\end{equation}
\indent The $\mathcal{T}_{\mathrm{HM}}$ integral can be readily approximated using the $N$-point trapezoidal rule,
\begin{align}
\mathcal{T}_{\mathrm{HM}}^{(N-1)} &\approx \frac{2}{c_{0}^{2}} \cdot \sum_{n=0}^{N-1} \frac{W_{\mathrm{HM}}\left(\tilde{t}_{n}, \vec{r}_{\mathrm{ph}}\left(\lambda_{n}\right)\right)+W_{\mathrm{HM}}\left(\tilde{t}_{n+1}, \vec{r}_{\mathrm{ph}}\left(\lambda_{n+1}\right)\right)}{2} \cdot\left(\tilde{t}_{n+1}-\tilde{t}_{n}\right)
\\
&= \frac{2 \cdot \varDelta t_{\mathrm{SR}}}{c_{0}^{2} \cdot N} \cdot \left( \sum_{n=1}^{N} W_{\mathrm{HM}}\left(\tilde{t}_{n}, \vec{r}_{\mathrm{ph}}\left(\lambda_{n}\right)\right) + \frac{W_{\mathrm{HM}}\left(\tilde{t}_{N}, \vec{r}_{\mathrm{ph}}\left(\lambda_{N}\right)\right) + W_{\mathrm{HM}}\left(\tilde{t}_{0}, \vec{r}_{\mathrm{ph}}\left(\lambda_{0}\right)\right)}{2} \right) \label{eqTHM}\\
\textrm{with time } \quad \tilde{t}_{n} & = t(\lambda_0) + \varDelta t_{\mathrm{SR}} \cdot \lambda_{n} = t(\lambda_0) + \varDelta t_{\mathrm{SR}} \cdot \frac{n}{N}, \qquad 0 \leq n \leq N,
\end{align}
with $(N-1)$ being the number of segments in the uniform grid sampling of the light path $\mathcal{P}$. Finally, the gravito-magnetic effect, the $\mathcal{T}_{\mathrm{SM}}$ term, can be approximated with a two-point trapezoidal rule as
\begin{equation}
\mathcal{T}_{\mathrm{SM}} \approx-\frac{2 GM_{e} R_{e}^{2}}{5 c_{0}^{3}} \cdot\left(\vec{\omega}_{e} \times \vec{r}_{e}\right) \vdp \vec{d}_{0} \cdot\left(\frac{1}{\left|\vec{r}_{e}\right|^{3}}+\frac{1}{\left|\vec{r}_{r}\right|^{3}}\right) \cdot \varDelta t_{\mathrm{SR}}. \label{eqTSM}
\end{equation}

Anticipating the result, it is beneficial to separate the special relativistic contribution into a delay $\Delta t_\textrm{inst}$ from the instantaneous inter-satellite range at the reception time $t_r$ and into a special relativistic correction $\mathcal{T}_\textrm{SR}$, i.e.~
\begin{align}
\varDelta t_{\mathrm{SR}} = \frac{\left|\vec{r}_{r}-\vec{r}_{e}\right|}{c_{0}} = \frac{\left|\vec{r}_{B}(t_r)-\vec{r}_{A}(t_e)\right|}{c_{0}} = \frac{\left|\vec{r}_{B}(t_r)-\vec{r}_{A}(t_r)\right|}{c_{0}} + \mathcal{T}_\textrm{SR} = \Delta t_\textrm{inst} + \mathcal{T}_\textrm{SR}, \label{eqTSR}
\end{align}
where it was assumed without loss of generality that the light is received by satellite $B$ after being emitted by satellite $A$ at time $t_e = t_r - \Delta t$.  In summary, the light propagation time $\Delta t_\textrm{rel}$ can be written as
\begin{equation}
{\varDelta t_{\mathrm{rel}}} = \varDelta t_{\mathrm{inst}} + \mathcal{T}
\end{equation}
with the light-time correction $\mathcal{T}$ containing special and general relativistic contributions
\begin{equation}
\mathcal{T} = \mathcal{T}_{\mathrm{SR}} + \mathcal{T}_{\mathrm{GR}} = \mathcal{T}_{\mathrm{SR}} + \mathcal{T}_{\mathrm{PM}} + \mathcal{T}_{\mathrm{HM}} + \mathcal{T}_{\mathrm{SM}}. 
\label{eqMathcalTsum}
\end{equation}
In order to compute all these terms, the emission position and emission time of the photon is needed, which depend on the light-time corrections. This yields an implicit light-time equation, which is solved in section~\ref{secLightTimeEquation}, after discussing the remaining correction for the atmosphere.

\section{Light time correction $\Delta t_\textrm{media}$ due to atmosphere}
\label{sectmedia}
At orbit heights below approx.~500\,km, such as the low Earth orbits of the GRACE and GRACE Follow-On satellites, the residual atmosphere may alter the speed of light due to refraction. A deviation of the refractive index $n$ from unity arises due to the neutral atmosphere and due to free electrons in the ionosphere. The former effect is negligible for interferometric range measurements, i.e.~for the time-delay $\Delta t_\textrm{media}$, since the fluctuations are estimated to be below $2~\textrm{nm}/\sqrt{\textrm{Hz}}/c_0$ for mHz frequencies, and with sinusoidal variations below 1\,nm/$c_0$ amplitude at once and twice the orbital frequency \cite{mueller2017design}.

However, the propagation of electromagnetic waves needs to be modeled according to propagation laws in plasma due to the charged particles in the ionosphere between $75..1000$\,km height. The main correction to the propagation time is the first-order ionospheric delay, which is commonly expressed as \cite{petit2010iers,Ionosphere}
\reviewermark{ \begin{align}
	\Delta t_\textrm{media} = \frac{1}{c_0} \int_{\mathcal{P}} (n-1)~\mathrm{d} s \approx - \frac{40.3\,\textrm{Hz}^2 /\textrm{m}}{c_0 \cdot f_\textrm{em}^2} \cdot \frac{\textrm{TEC}}{1\,e^-/\textrm{m}^2} = - \frac{40.3\,\textrm{Hz}^2}{f_\textrm{em}^2} \cdot \frac{\langle \eta \rangle }{1\,e^-/\textrm{m}^3}\cdot \Delta t_\textrm{SR}, \label{eqTmedia}
	\end{align} }
where \reviewermark{$f_\textrm{em}$} is the frequency of the electromagnetic wave and $\textrm{TEC}$ is the total electron content on the photon path with units of electrons per square-meter. The ionospheric delay is actually an advancement, since the correction is always negative, which is known from GNSS, where the code delay is positive, while the phase delay is negative. Due to the frequency dependence, it is possible to estimate variations of the TEC with interferometric range measurements at two different frequencies, but the absolute value of the TEC, and hence, the absolute value of $\Delta t_\textrm{media}$ is not measurable, because the ranging instruments can determine only a biased range.

However, in order to simulate the effect, the TEC can be expressed as the product of the mean electron density $\langle \eta \rangle$ between the satellites and the geometrical inter-satellite distance $\Delta t_\textrm{SR} \cdot c_0$. For satellites at a height of 400\,km, the electron density can reach values of up to $\langle \eta \rangle=10^{12}~e^{-}/\textrm{m}^3$ \cite{kelley2009earth}, which translate in worst-case to an absolute delay of $-13\,\textrm{mm}/c_0$ for a microwave frequency of $f = 24.5\,\textrm{GHz}$ and $\Delta t_\textrm{SR} \approx 200\, \textrm{km} /c_0$. \reviewermark{The effect of such a non-measurable absolute delay onto the instantaneous biased KBR range is assessed through the LTC in sec.~\ref{secLTCinDOWR}.
	On this occasion,} we point out that ionospheric effects are negligible for laser ranging with an optical frequency of $281\,\textrm{THz}$, since the contributions in propagation time or biased range are reduced by the factor
\begin{equation}
\left( \frac{24.5\,\textrm{GHz}}{281\,\textrm{THz}} \right)^2 \approx 7.6 \cdot 10^{-9}
\end{equation}
compared to the microwave K-band.

\section{Solving the light-time equation}
\label{secLightTimeEquation}
The propagation time $\Delta t$ of electromagnetic waves or photons between the two satellites has been described so far as a function of the photon path, or more precisely, as a function of the emission time $t_e$, emission position $\vec r_e$, reception time $t_r$ and reception position $\vec r_r$.

We may assume that the satellite trajectories are known, in particular, the satellite position $\vec r_{A/B}$, velocity $\dot{\vec r}_{A/B}$ and acceleration $\ddot{\vec r}_{A/B}$ at the time of reception $t_r$. The acceleration can be derived with a kinematic approach as time-derivative or by dynamic means using force models. Without loss of generality, we may assume that satellite $B$ is the receiver such that the reception position becomes $\vec r_r = \vec r_B(t_r)$ and that satellite $A$ is the emitter. 

Using Taylor expansion, the satellite's trajectory can be approximated in the vicinity of $t_r$ as
\begin{align}
\vec r_A(t_r - \epsilon) \approx \vec r_A(t_r) - \dot{\vec r}_A(t_r) \cdot \epsilon + \ddot{\vec r}_A(t_r) \cdot \epsilon^2/2, \label{eqTaylorSC}
\end{align}
which allows us to write the position at the event of photon emission as $ \vec r_e = \vec r_A(t_r - \Delta t)$. In order to solve for $\Delta t$ one has to solve the implicit equation
\begin{align}
\Delta t(t_r) &= \frac{| \vec r_B(t_r) - \vec r_A(t_r - \Delta t) |}{c_0} + \mathcal{T}_\textrm{GR}(\vec r_e = \vec r_A(t_r - \Delta t)) + \Delta t_\textrm{media}(\vec r_e = \vec r_A(t_r - \Delta t))
\label{eqFullLightConeEq1}
\end{align}

A solution can be obtained by iterative means using
\begin{align}
\Delta t^{(n+1)}(t_r) &= \frac{| \vec r_B(t_r) - \vec r_A(t_r - \Delta t^{(n)}) |}{c_0} \nonumber \\ 
&\quad+ \mathcal{T}_\textrm{GR}(\vec r_e = \vec r_A(t_r - \Delta t^{(n)})) + \Delta t_\textrm{media}(\vec r_e = \vec r_A(t_r - \Delta t^{(n)}))
\label{eqFullLightConeEq2}
\end{align}
with start value $\Delta t^{(0)} = \Delta t_\textrm{inst} = |\vec r_A(t_r) -\vec r_B(t_r)|/c_0$. The three summands on the right hand side have an amplitude of approximately 200\,km/$c_0$, 300\,$\mu$m/$c_0$ and in case of the K-band -13\,mm/$c_0$, respectively.

The vectors in the first term have typically a magnitude of $7 \cdot 10^6$~meters, which limits direct numerical solutions of  \reviewermark{$\vec r_A - \vec r_B$ in} eq.~(\ref{eqFullLightConeEq2}) to a precision of the order of nanometer/$c_0$ due to the $\approx$\,15 digits precision of \reviewermark{64-bit (double) floating-point arithmetic. One way to overcome this limitation is to derive an analytical closed-form solution for $\Delta t$.  This can be achieved by substituting eq.~(\ref{eqTaylorSC}) into eq.~(\ref{eqFullLightConeEq2}), taking into account the relation in eq.~(\ref{eqTSR}),  and evaluating the first few iterations using an algebraic manipulation software. If terms with negligible magnitude are omitted in the lengthy expression\footnote{\reviewermark{ We evaluated all individual terms using GRACE-FO orbit data and omitted terms with a magnitude $10^{-12}\,\textrm{m}/c_0$. One can reproduce our set of relevant terms by using the book-keeping parameter $\epsilon^n$ and exploiting the replacement rules: $c_0 \rightarrow c_0 \cdot \epsilon^{-2}$, $\Delta t_\textrm{inst} \rightarrow \Delta t_\textrm{inst} \cdot \epsilon^1$, $(\mathcal{T}_\textrm{GR}+\Delta t_\textrm{media}) \rightarrow (\mathcal{T}_\textrm{GR}+\Delta t_\textrm{media}) \cdot \epsilon^4$, $\dot{\vec{r}}_A \vdp \dot{\vec{r}}_A \rightarrow (\dot{\vec{r}}_A \vdp \dot{\vec{r}}_A) \cdot \epsilon^{-1}$. Eq.~(\ref{eqOWRanaPre})-(\ref{eqOWRanaGR}) is a series expansion up to order $\epsilon^{6}$ of eq.~(\ref{eqFullLightConeEq2}) for $\Delta t^{(2)}$. Using this expansion or threshold magnitude, the result does not change for higher iteration numbers. }}, the solution for $\Delta t^{(2)}$ (and higher iteration numbers) reads}
\begin{align}
\Delta t(t_r) &= \Delta t_\textrm{inst}(t_r) + \mathcal{T}_\textrm{SR}(t_r) + \mathcal{T}_\textrm{GR}(t_r) + \Delta t_\textrm{media}(t_r) \label{eqOWRanaPre}
\\
\mathcal{T}_\textrm{SR} &= \Delta t_\textrm{inst} \frac{\vec d_0 \vdp \dot{\vec{r}}_A }{c_0}  + \Delta t_\textrm{inst}^2  \frac{\vec d_0 \vdp \ddot{\vec{r}}_A }{2 \cdot c_0} + \frac{ \Delta t_\textrm{inst}^2 \cdot (- \vec{d}_0 \vdp \ddot{\vec{r}}_A \cdot \vec{d}_0 \vdp \dot{\vec{r}}_A -
	\dot{\vec{r}}_A \vdp \ddot{\vec{ r}}_A/2) + \Delta t_\textrm{inst}/2 \cdot ( (\vec{d}_0 \vdp \dot{\vec{r}}_A)^2 +  |\dot{\vec{r}}_A|^2 )}{c_0^2} \nonumber \\
&\quad + \frac{\Delta t_\textrm{inst} \cdot \vec{d}_0 \vdp \dot{\vec{r}}_A \cdot |\dot{\vec{r}}_A|^2  }{c_0^3} + (\mathcal{T}_\textrm{GR}+\Delta t_\textrm{media}) \frac{\vec d_0 \vdp \dot{\vec{r}}_A}{c_0} +  \mathcal{O} \left( 10^{-12}\,\textrm{m}/c_0 \right) \label{eqOWRana}\\
\mathcal{T}_\textrm{GR} &= \mathcal{T}_\textrm{GR}(\vec r_e = \vec r_A(t-\Delta t_\textrm{inst} - \Delta t_\textrm{inst} \cdot \vec d_0 \vdp \dot{\vec{r}}_A / c_0 )) \approx \mathcal{T}_\textrm{GR}(\vec r_e = \vec r_A(t-\Delta t)) \label{eqOWRanaGR}
\end{align}
where all quantities, also the one used for calculating $\vec d_0$ with eq.~(\ref{eqd0def}), are evaluated at the photon reception time $t_r$. Thus, the equation can be directly applied with orbit data from GRACE or GRACE Follow-On.

The overall light-time correction $\mathcal{T} = \mathcal{T}_\textrm{SR} + \mathcal{T}_\textrm{GR}$ is dominated by the first term in eq.~(\ref{eqOWRana}), which has an amplitude of the order of $-5$\,m/$c_0$ for $\Delta t_\textrm{inst} \approx 200\,\textrm{km}/c_0$ and $\vec d_0 \vdp \dot{\vec{r}}_A  \approx -7.6\,\textrm{km/s}$. The derivation of $\mathcal{T}$ assumed so far a single path of a photon from one satellite to the other, i.e.~an one-way ranging approach. However, the ranging systems in GRACE and GRACE-Follow-On exchange light in both directions and the light-time correction becomes a linear combination of two (LRI) or four (MWI) one-way corrections ($\mathcal{T}$). As will turn out subsequently, these linear combinations have a significantly lower magnitude due to a high common-mode rejection.

\section{Light time correction in dual one-way ranging (DOWR)}
\label{secLTCinDOWR}
\begin{figure}
	\centering
	\includegraphics[scale=0.5]{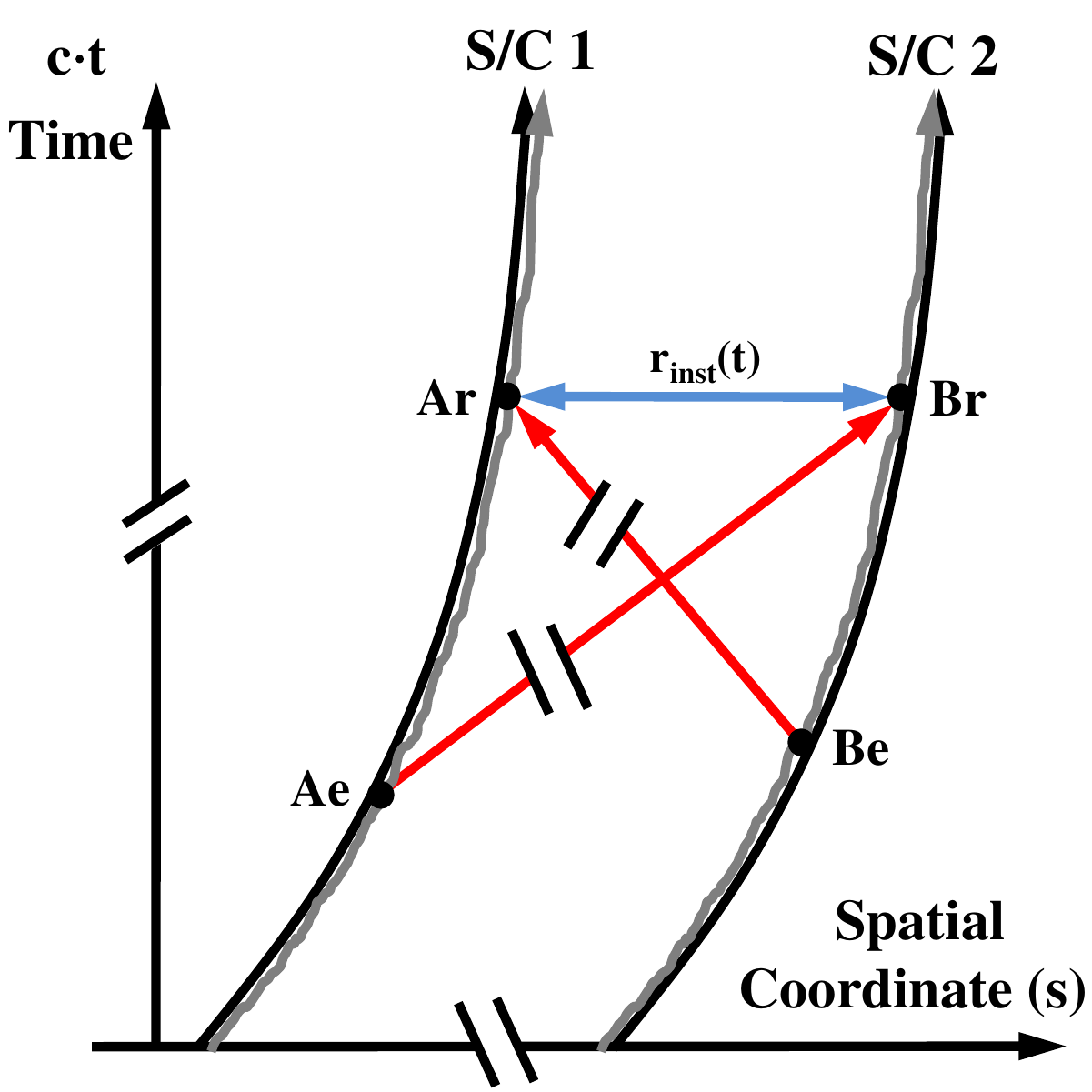}
	\qquad
	\includegraphics[scale=0.5]{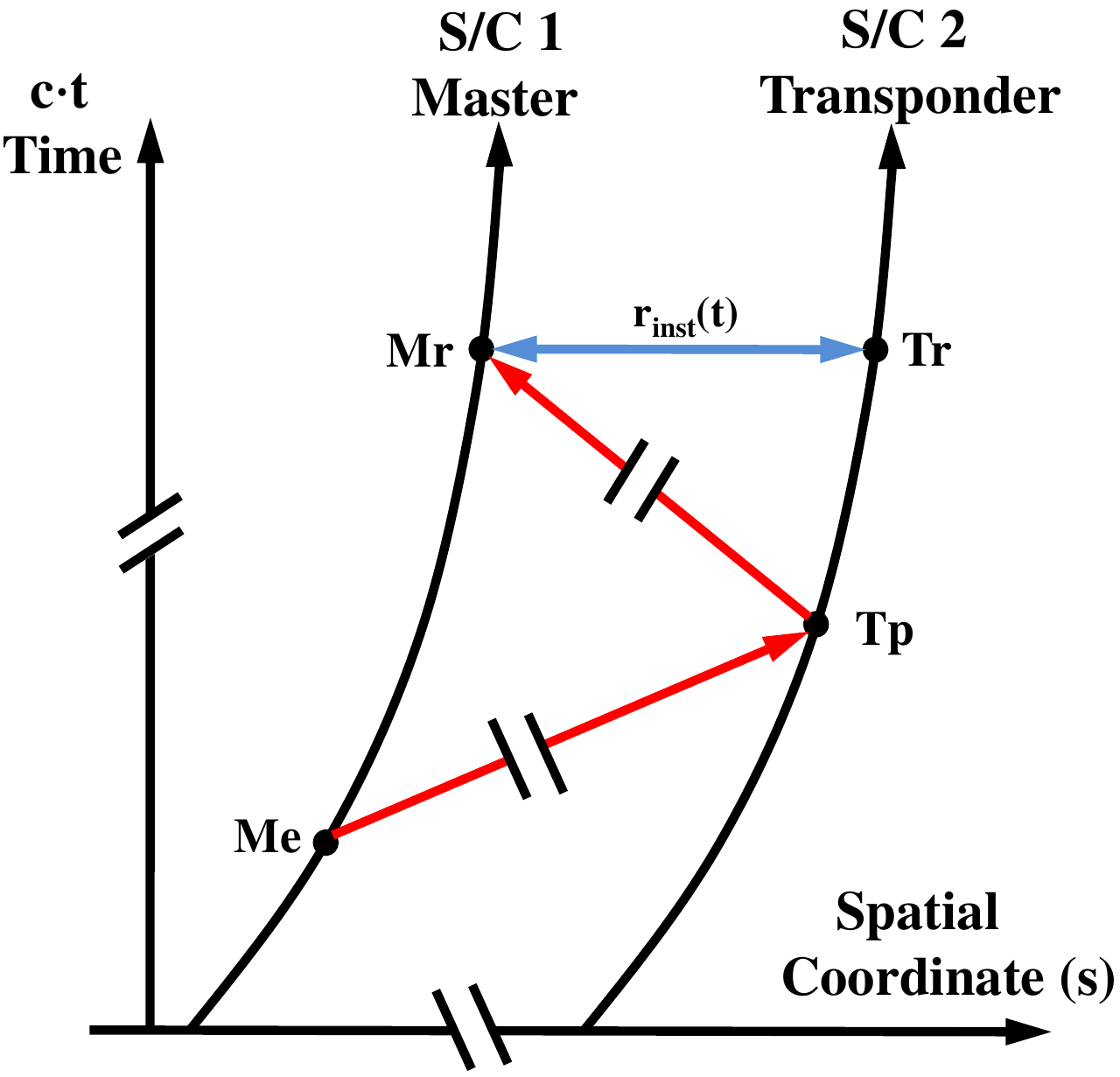}
	\caption{Minkowski diagram of the light path (red arrows) in a dual-one way ranging (DOWR) scheme at a particular frequency (left plot) and in the two-way ranging (TWR) scheme (right plot). For the DOWR, the emission (e) and reception (r) events are located at the antenna phase centers (grey trajectories) of the two satelltes (A and B). In the TWR case, these events occur at the center-of-mass (solid black lines) of the master (M) and transponder (T) satellite. The reflection event on the transponder side is denoted as Tp. }
	\label{fig:Minkowski}
\end{figure}

The dual-one way ranging concept is used by the microwave ranging systems in GRACE and GRACE Follow-On \cite{wen_nasa_nodate}, where the ionospheric effect needs to be removed using measurements at two frequencies, namely at the K-band with 24.5\,GHz and at the Ka-band with 32.7\,GHz frequency. Each satellite ($A$ and $B$) records two phase measurements ($\Phi_A^{K}$, $\Phi_A^{Ka}$, $\Phi_B^{K}$ and $\Phi_B^{Ka}$) using heterodyne interferometry and phase tracking, which represent the phase difference between a local (LO) and a received (RX) electromagnetic field at reception time $t_r$,~i.e.~\cite[eq.~2.14]{kim2000simulation}

\reviewermark{
	\begin{align}
	\Phi_{B}^{K/Ka}(t_r) = \Phi_{Br}^{K/Ka} &= -\left( \varphi_\textrm{RX,B} - \varphi_\textrm{LO,B} \right) 
= - \left( \reviewermark{\hat{f}}_{A}^{K/Ka} \cdot \tau_A^\reviewermark{\textrm{USO}}(t_r-\Delta t_{AeBr}^{K/Ka}) - \reviewermark{\hat{f}}_{B}^{K/Ka} \cdot \tau_B^\reviewermark{\textrm{USO}}(t_r) \right) \label{eqRawDowrSatellitePhaseA1}\\ 
&\approx -\left(\reviewermark{\hat{f}}_{A}^{K/Ka} \cdot \tau_A^\reviewermark{\textrm{USO}}(t_r) - \reviewermark{\hat{f}}_{B}^{K/Ka} \cdot \tau_B^\reviewermark{\textrm{USO}}(t_r)\right) + \reviewermark{\hat{f}}_{A}^{K/Ka} \cdot \frac{\textrm{d} \tau_A^\reviewermark{\textrm{USO}}}{\textrm{d}t}\cdot \Delta t_{AeBr}^{K/Ka} + \textrm{const.}  \label{eqRawDowrSatellitePhaseA2} \\
&= -\left(\reviewermark{\hat{f}}_{A}^{K/Ka} \cdot \tau_A^\reviewermark{\textrm{USO}}(t_r) - \reviewermark{\hat{f}}_{B}^{K/Ka} \cdot \tau_B^\reviewermark{\textrm{USO}}(t_r)\right) + f_{A}^{K/Ka}(t_r) \cdot \Delta t_{AeBr}^{K/Ka} + \textrm{const.} \label{eqRawDowrSatellitePhaseA3} \\
\Phi_{A}^{K/Ka}(t_r) = \Phi_{Ar}^{K/Ka} &= +\left( \varphi_\textrm{RX,A} - \varphi_\textrm{LO,A} \right)  
= + \left( \reviewermark{\hat{f}}_{B}^{K/Ka} \cdot \tau_B^\reviewermark{\textrm{USO}}(t_r-\Delta t_{BeAr}^{K/Ka}) - \reviewermark{\hat{f}}_{A}^{K/Ka} \cdot \tau_A^\reviewermark{\textrm{USO}}(t_r) \right)  \label{eqRawDowrSatellitePhaseB1}\\ 
&\approx +\left(\reviewermark{\hat{f}}_{B}^{K/Ka} \cdot \tau_B^\reviewermark{\textrm{USO}}(t_r) - \reviewermark{\hat{f}}_{A}^{K/Ka} \cdot \tau_A^\reviewermark{\textrm{USO}}(t_r)\right) - \reviewermark{\hat{f}}_{B}^{K/Ka} \cdot \frac{\textrm{d} \tau_B^\reviewermark{\textrm{USO}}}{\textrm{d}t}\cdot \Delta t_{BeAr}^{K/Ka} + \textrm{const.}  \label{eqRawDowrSatellitePhaseB2} \\
&= +\left(\reviewermark{\hat{f}}_{B}^{K/Ka} \cdot \tau_B^\reviewermark{\textrm{USO}}(t_r) - \reviewermark{\hat{f}}_{A}^{K/Ka} \cdot \tau_A^\reviewermark{\textrm{USO}}(t_r)\right) - f_{B}^{K/Ka}(t_r) \cdot \Delta t_{BeAr}^{K/Ka} + \textrm{const.} \label{eqRawDowrSatellitePhaseB3}
	\end{align}
}

	The phases $\varphi_\textrm{...}$ of the electro-magnetic fields are given as the product of a \reviewermark{static nominal} \reviewersout{proper} frequency $\reviewermark{\hat{f}}_{A/B}^{K/Ka}$  and \reviewersout{proper} \reviewermark{USO time $\tau_{A/B}^\textrm{USO}$, which differs from the proper time $\tau_{A/B}$ due to clock errors. These clock errors account for noise and errors sources, in particular for deviations of the USO frequency} from the nominal or design values: $\hat{f}_{A}^{K} = 5076 \cdot 4.832\,\textrm{MHz}$, $\hat{f}_{A}^{Ka} = 6768 \cdot 4.832\,\textrm{MHz}$, $\hat{f}_{B}^{K} = 5076 \cdot 4.832099\,\textrm{MHz}$ and $\hat{f}_{B}^{Ka} = 6768 \cdot 4.832099\,\textrm{MHz}$ \cite{wen_nasa_nodate}. \reviewermark{The clock errors can be estimated} during precise orbit determination (see CLK1B and USO1B data products in GRACE-FO) and allow to derive the apparent frequencies $ f_{A/B}(t) = \reviewermark{\hat{f}}_{A/B} \cdot  \textrm{d} \tau_{A/B}^\reviewermark{\textrm{USO}} / \textrm{d}t$, which \reviewermark{are relevant for the ranging and contain effects from } relativistic time dilation \reviewermark{and clock errors, e.g.~USO frequency deviations}. For the purpose of calculating the light-time-correction, which is significantly smaller than the actual ranging signal, it is usually sufficient to drop the time-dependency and use a (daily) mean value $\langle f_{A/B} \rangle$, since the deviations of $ f_{A/B}(t)/\langle f_{A/B} \rangle$ \reviewermark{from unity} are below $10^{-10}$ in magnitude for \reviewermark{both, the daily clock drifts and } the relativistic modulation\footnote{ A typical spectrum of the proper time $\tau(t)$ for a GRACE-like satellite is shown in \cite[Fig.~2.14]{mueller2017design}, which has a dominant peak with a rms-amplitude of approx.~$10^{-7}~\textrm{s}/\sqrt{\textrm{Hz}}$ at the orbital frequency ($\approx 0.18$\,mHz). Using the provided equivalent noise bandwidth of $ 24\,\mu$Hz, one can convert the value to an amplitude for $\textrm{d}\tau/\textrm{d}t$, i.e.~$10^{-7}~\textrm{s}/\sqrt{\textrm{Hz}} \cdot \sqrt{24\,\mu\textrm{Hz}} \cdot (2 \pi \cdot \textrm{0.18\,mHz}) \cdot \sqrt{2} \approx 10^{-12}$. }.
	
	The first part of $\Phi$ in line (\ref{eqRawDowrSatellitePhaseA3}) and (\ref{eqRawDowrSatellitePhaseB3}) is proportional to $\hat{f}_B \cdot \tau_B^\reviewermark{\textrm{USO}} (t_r)- \hat{f}_A \cdot \tau_A^\reviewermark{\textrm{USO}}(t_r)$ and describes a constant positive phase ramp with a slope of approx.~500\,kHz and 670\,kHz for the K- and Ka-band, respectively. The frequency order is reversed between the spacecraft.  Usually, phase trackers are not aware of the frequency order and return a positive slope, which means that the sign of the second term ($\Delta t$) is reversed between both S/C. This sign convention is consistent with the usual description of phase-tracking in the laser ranging instrument (see next section). However, it is opposite to the usual literature for microwave ranging (see \cite{wen_nasa_nodate}), where the phase ramps on  satellite~A (GFO-C) have negative slope. The term $\Delta t_{AeBr}$ in above eq.~describes the propagation time of the microwaves from satellite $A$ to $B$, while $\Delta t_{BeAr}$ denotes the opposite path. The last summand \textit{const.} represents the fact that the phase measurement always have an unknown bias, which is constant unless the phase-tracking is interrupted or cycle slips occur. The MWI measures distance variations between the antenna phase center (APC), which are offset on each satellite by approx.~1.4\,m in the direction of the distant satellite.
	
By subtracting the two phase observations in the K- or Ka-band, and dividing with the sum of the \reviewersout{nominal (or measured)} \reviewermark{measured apparent frequencies $f_{A/B,\textrm{meas}}^{K/Ka}$} (cf. the eq.~2.16 in \cite{kim2000simulation}), one can obtain a range observation at the K- and Ka-band, i.e
\begin{align}
\rho_\textrm{DOWR}^{K/Ka}(t) &= c_0 \cdot \int^t_0 \frac{ \textrm{d}~\left(\Phi_{Br}^{K/Ka}(t^\prime) - \Phi_{Ar}^{K/Ka}(t^\prime) \right)/ \textrm{d}t^\prime}{f_{A,\textrm{meas}}^{K/Ka}(t^\prime) + f_{B,\textrm{meas}}^{K/Ka}(t^\prime)} \textrm{d}t^\prime \label{eqrhoDOWRKKa2pre}\\
&\approx c_0 \cdot \frac{ \Phi_{Br}^{K/Ka} - \Phi_{Ar}^{K/Ka}}{\langle f_{A,\textrm{meas}}^{K/Ka}\rangle + \langle f_{B,\textrm{meas}}^{K/Ka} \rangle} = c_0 \cdot \frac{ f_{A}^{K/Ka}(t) \cdot \Delta t_{AeBr}^{K/Ka} +  f_{B}^{K/Ka}(t) \cdot \Delta t_{BeAr}^{K/Ka}}{ \langle f_{A,\textrm{meas}}^{K/Ka} \rangle + \langle f_{B,\textrm{meas}}^{K/Ka}\rangle} +  \textrm{const.}  \label{eqrhoDOWRKKa2pre2}\\
&\approx c_0 \cdot \Delta t_\textrm{inst,APC} + c_0 \cdot \frac{ \langle f_A^{K/Ka} \rangle \cdot \mathcal{T}_{AeBr}^{K/Ka} + \langle f_B^{K/Ka} \rangle \cdot \mathcal{T}_{BeAr}^{K/Ka}}{\langle f_A^{K/Ka} \rangle + \langle f_B^{K/Ka} \rangle }   \nonumber \\ &\quad + c_0 \cdot \frac{ \langle f_A^{K/Ka} \rangle \cdot \Delta t_\textrm{media}^{K/Ka} + \langle f_B^{K/Ka} \rangle \cdot \Delta t_\textrm{media}^{K/Ka}}{ \langle f_A^{K/Ka} \rangle + \langle f_B^{K/Ka}\rangle} +  \textrm{const.} \\
&= \rho_\textrm{inst,APC} + c_0 \cdot \mathcal{T}_\textrm{DOWR}^{K/Ka} + \rho_\textrm{media}^{K/Ka}  +  \textrm{const.},\label{eqrhoDOWRKKa2}
\end{align}
which can be written as the sum of instantaneous distance between APC $\rho_\textrm{inst,APC}$, light time effect $\mathcal{T}_\textrm{DOWR}^{K/Ka}$ and ionospheric delay $\rho_\textrm{media}^{K/Ka}$. The light paths in the DOWR scheme  are shown for a single frequency in the left plot of fig.~\ref{fig:Minkowski}. \reviewermark{Eq.~(\ref{eqrhoDOWRKKa2pre}) is suited to convert the measured phases to the DOWR ranges $\rho_\textrm{DOWR}^{K/Ka}$. For the derivation of the much smaller light-time and ionospheric corrections, the approximations in eq.~(\ref{eqrhoDOWRKKa2pre2})-(\ref{eqrhoDOWRKKa2}) are usually sufficient, where the distinction between true apparent and measured apparent frequency, as well as their time-dependencies, are omitted. }

One can remove the ionospheric effect by a linear combination of $\rho_\textrm{DOWR}^{K}$ and $\rho_\textrm{DOWR}^{Ka}$, which yields the DOWR biased range as
\begin{align}
\rho_\textrm{DOWR} &=  a^{Ka} \cdot \rho_\textrm{DOWR}^{Ka} + a^{K} \cdot \rho_\textrm{DOWR}^{K} = \rho_\textrm{inst,APC} + c_0 \cdot  \mathcal{T}_\textrm{DOWR} +  \textrm{const.} \label{eqDOWReq3}
\end{align}
where the light-time effect $\mathcal{T}_\textrm{DOWR}$ is, in general, a function of four $\mathcal{T}^{K/Ka}_{...}$ terms arising from two photon paths at two frequencies:
\begin{align}
\mathcal{T}_\textrm{DOWR} &= a^\textrm{K} \cdot \mathcal{T}_\textrm{DOWR}^{K} + a^\textrm{Ka} \cdot \mathcal{T}_\textrm{DOWR}^{Ka} \\
& =   b_{AeBr}^{K} \cdot \mathcal{T}_{AeBr}^K + b_{AeBr}^{Ka} \cdot \mathcal{T}_{AeBr}^{Ka} + b_{BeAr}^{K} \cdot \mathcal{T}_{BeAr}^{K} + b_{BeAr}^{Ka} \cdot \mathcal{T}_{BeAr}^{Ka} 
\end{align}
with $a^{K/Ka}_{...}$ and $b^{K/Ka}_{...}$ coefficients given in table~\ref{tab:MWIcoeff}. 

The biased dual-one way range $\rho_\textrm{DOWR}$ is apportioned in eq.~(\ref{eqDOWReq3}) into the instantaneous range $\rho_\textrm{inst,APC}$ and an effect due to the finite speed of light $c_0 \cdot \mathcal{T}_\textrm{DOWR}$. In order to obtain the instantaneous range, one has to remove this light-time effect using an estimate or correction $\widehat{\mathcal{T}}_\textrm{DOWR}$, which can be derived from orbit data. Moreover, an antenna offset correction is applied in order to transform the biased range between APC into a biased range between the center-of-mass that is usually used for gravity field recovery.

The cross coupling of $\Delta t_\textrm{media}$ into the light-time correction $\mathcal{T}_\textrm{DOWR}$ is usually omitted (cf.~eq.~(\ref{eqOWRana})), i.e.~the K and Ka superscripts of $\mathcal{T}$ are dropped
\begin{align}
\widehat{\mathcal{T}}_\textrm{DOWR} \approx  b_{AeBr} \cdot \mathcal{T}_{AeBr} + b_{BeAr}  \cdot \mathcal{T}_{BeAr}, \label{eqTDOWRcombin}
\end{align}
because the absolute value of the ionospheric delay $\Delta t_\textrm{media}$ is difficult to estimate and the effect on the final correction $\mathcal{T}_\textrm{DOWR}$ is well below the microwave instrument resolution. In other words, the LTC computation neglects any atmospheric effect, i.e.~the photons at K- and Ka-Band have the same emission time as in vacuum and travel along the same path. However, this approximation does not affect the phase delay as determined and corrected for with the ionospheric correction (cf.~sec.~(\ref{sectmedia})). The omission error in the LTC is at the sub-picometer level and can be assessed using eq.~(\ref{eqTmedia}) and eq.~(\ref{eqOWRana}), i.e.~
\begin{align}
\left| c_0 \mathcal{T}_\textrm{DOWR,media} \right| &= \left| \frac{40.3\,\textrm{Hz}^2}{c_0} \cdot \frac{\textrm{TEC}}{1\,e^-/\textrm{m}^3}  \cdot \left( \frac{ b_{AeBr}^{K} \cdot \vec d_0 \vdp \dot{\vec{r}}_A}{(f_A^K)^2} - \frac{ b_{BeAr}^{K} \cdot  \vec d_0 \vdp \dot{\vec{r}}_B}{(f_B^K)^2} +  \frac{b_{AeBr}^{Ka} \cdot \vec d_0 \vdp \dot{\vec{r}}_A}{(f_A^{Ka})^2} -  \frac{b_{BeAr}^{{Ka}} \cdot  \vec d_0 \vdp \dot{\vec{r}}_B}{(f_B^{Ka})^2} \right) \right|  \\
&\approx \left| -2 \cdot 10^{-13}\,\textrm{m} - 8 \cdot 10^{-18}\,\textrm{m} \cdot \frac{\dot \rho_\textrm{inst}}{1\,\textrm{m/s}} \right| < 10^{-12}\,\textrm{m}
\end{align}
where $\vec d_0 = (\vec r_B-\vec r_A)/|\vec r_B-\vec r_A| $, $ \vec d_0 \vdp \dot{\vec{r}}_A = -7.6\,\textrm{km/s}$, and $ \vec d_0 \vdp \dot{\vec{r}}_B = 7.6\,\textrm{km/s}+\reviewermark{\dot \rho_\textrm{inst}}$ were used as values. The range rate \reviewermark{$\dot \rho_\textrm{inst}$}\reviewersout{$\dot \rho$} is usually below 1\,m/s, hence, the modulation due to $\rho_\textrm{inst}$  is  insignificant. The same holds for variations of the TEC, which can be expected to be well below the used upper bound estimate $\textrm{TEC} = 10^{12}\,\textrm{e}^-/\textrm{m}^3 \cdot 200\,\textrm{km} $.

The \reviewermark{leading terms} of the DOWR light-time correction in the range domain, which has to be subtracted from the measured biased range $\rho_\textrm{DOWR}$ to obtain the instantaneous range, reads \reviewermark{
	\begin{align}
	c_0 \widehat{\mathcal{T}}_\textrm{DOWR} &= \Delta t_\textrm{inst} \cdot \left(  b_{AeBr} \cdot \vec d_0 \vdp \dot{\vec{r}}_A-b_{BeAr} \cdot  \vec d_0 \vdp \dot{\vec{r}}_B \right) + \textrm{const.} + \ldots
	= -\frac{ |\vec{r}_B-\vec{r}_A| \cdot \dot{\rho}_\textrm{inst,OD}}{2 \cdot c_0} + \textrm{const.} + \ldots, \label{eqDOWRdominating}
	\end{align}}
where \reviewermark{both shown} terms have a typical magnitude of a few hundred micrometers \reviewermark{(cf.~table~\ref{tab:GReffects}). The $\dot{\rho}_\textrm{inst,OD}$ denotes the instantaneous range rate from orbit data (OD). This leading term describes approximately 99.9\,\% of the LTC at once and twice the orbit frequency, which may be sufficient in some cases. However, the analyses in this paper consider the full expression, not just the leading term.  }

\reviewersout{ The last term proportional to $\Delta b = b_{AeBr} - b_{BeAr} \approx \Delta f^{K}/(2f^{K}) \approx -10^{-5}$ depends on the fractional frequency difference between satellites $\Delta f^{K} = f_{A}^{K}-f_{B}^{K}$ and is scaled by the absolute satellite velocity along the line-of-sight ($\approx$\,7600\,m/s).}

\begin{table}
	\centering
	\caption{Numerical values for coefficients introduced to describe the light time correction in dual one-way ranging, which are based on carrier frequencies in the K and Ka band for the microwave ranging system.}
	\label{tab:MWIcoeff}
	\renewcommand{\arraystretch}{1.5}
	\begin{tabular}{lll}
		\hline\noalign{\smallskip}
		Name & Formula & Nominal Value \reviewermark{($ f = \hat{f}$)}  \\
		\noalign{\smallskip}\hline\noalign{\smallskip}
		$a^K$   & $-f_A^{K} \cdot f_B^{K}/(f_A^{Ka} \cdot f_B^{Ka}-f_A^{K} \cdot f_B^{K})$ & -9/7   \\
		$a^{Ka}$  & $f_A^{Ka} \cdot f_B^{Ka}/(f_A^{Ka} \cdot f_B^{Ka}-f_A^{K} \cdot f_B^{K})$ &  16/7  \\
		$b_{AeBr}^K$  & $\frac{(f_A^K)^2 \cdot f_B^K}{(f_A^K+f_B^K) (f_A^K f_B^K - f_A^{Ka} f_B^{Ka})}$ & $\frac{-43488000}{67648693}\approx -0.642851 $  \\
		$b_{AeBr}^{Ka}$ & $- \frac{(f_A^{Ka})^2 \cdot f_B^{Ka}}{(f_A^{Ka}+f_B^{Ka}) (f_A^K f_B^K - f_A^{Ka} f_B^{Ka})}$ &  $\frac{77312000}{67648693}\approx 1.1428454 $ \\
		$b_{BeAr}^{K}$  & $\frac{f_A^K \cdot (f_B^K)^2}{(f_A^K+f_B^K) (f_A^K f_B^K - f_A^{Ka} f_B^{Ka})}$ &  $\frac{-43488891}{67648693}\approx -0.642864 $ \\
		$b_{BeAr}^{Ka}$ & $- \frac{f_A^{Ka} \cdot (f_B^{Ka})^2}{(f_A^{Ka}+f_B^{Ka}) (f_A^K f_B^K - f_A^{Ka} f_B^{Ka})}$ & $\frac{77313584}{67648693}\approx 1.142869 $  \\
		$b_{AeBr}$  & $b_{AeBr}^{K}+b_{AeBr}^{Ka}$ &  $\approx 0.499995$ \\ 
		$b_{BeAr}$  & $b_{BeAr}^{K}+b_{BeAr}^{Ka}$ &  $\approx 0.500005$ \\  
		\noalign{\smallskip}\hline
	\end{tabular}
\end{table}

\section{Light time correction in two-way ranging (TWR)}
\label{secLTCinTWR}
\reviewermark{
The laser ranging instrument aboard GRACE-Follow-On is based on a master-transponder scheme, which is also called a two way ranging scheme. The role of master and transponder is inter-changeable between the satellites. As shown on the right plot in figure~\ref{fig:Minkowski}, the master satellite emits a photon at event $M_e$ using a frequency-stabilized laser source. The optical phase \reviewermark{(in cycles)} of this photon can be modelled as a function of the coordinate time $t$
		\begin{align}
\reviewermark{\varphi_\textrm{M}(t) = \int_{0}^{t} \tilde{f}_M(t^\prime)\cdot \frac{\textrm{d}\tau_M(t^\prime)}{\textrm{d}t^\prime} ~\textrm{d}t^\prime 
}
\end{align}
\reviewermark{where $\tilde{f}_M$ is the instantaneous optical laser frequency that would be measured in a rest-frame at the laser source and $\tau_M$ refers to the proper time of the master satellite. Imperfections of the laser or cavity, i.e.~frequency variations, can be accounted for by the time-dependent $\tilde{f}_M$}. 

	The photon emitted by the master satellite propagates to the transponder craft. The transponder utilizes a frequency-locked loop with 10\,MHz frequency offset. This means the laser phase $\varphi_\textrm{LO,T}(t)$, more precisely the time-derivative of it, is controlled such that the beatnote phase $\Phi_{T}(t)$, given as the phase difference between received (RX) and local oscillator (LO) light,  becomes
	\begin{align}
	\Phi_{T}(t) = \varphi_\textrm{LO,T}-\varphi_\textrm{RX,T} = \varphi_\textrm{LO,T}(t)-\varphi_\textrm{M}(t-\Delta t_\textrm{MeTp}(t)) =  +10\,\textrm{MHz} \cdot \tau_T^\textrm{USO}(t) + \varphi_\epsilon(t) + \textrm{const.}
	\label{eqMasterPhaseT}
	\end{align}
		\reviewermark{where $\tau_M^\textrm{USO}$ is the time of the ultra-stable oscillator clock, which may differ from the proper time $\tau_M$ due to noise or errors sources.
	The beatnote phase $\Phi_{T}$} implies that the optical phase of the transponder laser with units of cycles is
	\begin{align}
	\varphi_\textrm{LO,T}(t) = \varphi_\textrm{M}(t-\Delta t_\textrm{MeTp}(t)) + 10\,\textrm{MHz} \cdot \tau_T^\textrm{USO}(t) + \varphi_\epsilon(t) + \textrm{const.},
	\end{align}
	where $\varphi_\epsilon(t)$ was used to account for phase-variations that were not fully suppressed by the feedback control loop, e.g.~due to finite gain and bandwidth. These are much smaller than the phase ramp with a slope of 10\,MHz. The loop ensures a constant phase relation between emitted and received light on the transponder side, in other words, the transponder seems to reflect the received light at event $T_p$, however, with enhanced light power and slightly different frequency.
	
	Eventually, the transponder photon returns to the master side at the reception event $M_r$. The phase of the beatnote on the master satellite $\Phi_{M}$ reads
		\begin{align}
\Phi_{M}(t_r) &= \varphi_\textrm{RX,M} - \varphi_\textrm{LO,M} = \varphi_\textrm{LO,T}(t_r-\Delta t_\textrm{TpMr}) - \varphi_M(t_r) \\
&= \varphi_\textrm{M}(t_r-\Delta t_\textrm{TpMr}-\Delta t_\textrm{MeTp})  - \varphi_M(t_r) + 10\,\textrm{MHz} \cdot \tau_T^\textrm{USO}(t_r-\Delta t_\textrm{TpMr}) + \varphi_\epsilon(t_r-\Delta t_\textrm{TpMr})  + \textrm{const.} \\
&\approx -\frac{d \varphi_\textrm{M}}{d \tau_M} \cdot \frac{d \tau_M}{d t} \cdot (\Delta t_\textrm{TpMr}+\Delta t_\textrm{MeTp}) + 10\,\textrm{MHz} \cdot \tau_T^\textrm{USO}(t_r-\Delta t_\textrm{TpMr}) + \varphi_\epsilon(t_r-\Delta t_\textrm{TpMr}) + \textrm{const.} \\
& = -f_M(t_r) \cdot (\Delta t_\textrm{TpMr}+\Delta t_\textrm{MeTp}) + 10\,\textrm{MHz} \cdot \tau_T^\textrm{USO}(t_r-\Delta t_\textrm{TpMr}) + \varphi_\epsilon(t_r-\Delta t_\textrm{TpMr}) + \textrm{const.} \label{eqMasterPhaseM}
\end{align}
	The ranging information is encoded in the term containing the product of true apparent optical frequency ($f_M = \tilde{f}_M \cdot \textrm{d}\tau_M/\textrm{d}t$ ) and photon time of flight $\Delta t_{...}$. It can give rise to Doppler shifts of up to a few \,MHz over one orbital revolution.
	
	Subtracting both phase observations, when the transponder phase is temporally aligned to the master using an estimated one-way light travel time $\Delta t_\textrm{TpMr,est}$, removes the 10\,MHz phase ramp and the phase residuals $\varphi_\epsilon$.  Then, the phase difference is converted to a biased range observable using an estimate of the apparent optical frequency\footnote{\reviewermark{The LRI optical frequency $f_{M,\textrm{est}}(t)$, i.e.~the scale factor, is determined on a daily basis by comparing LRI and MWI range in the official GRACE-FO RL04 dataset.}} $f_{M,\textrm{est}}(t)$, as in the DOWR case (cf.~eq.~(\ref{eqrhoDOWRKKa2pre})), i.e.
	\begin{align}
		\rho_\textrm{TWR}(t) &= c_0 \cdot \int_0^t \frac{ \textrm{d} \left( \Phi_{T}(t^\prime-\Delta t_\textrm{TpMr,est}) - \Phi_{M}(t^\prime) \right) / \textrm{d}t^\prime}{2 \cdot f_{M,\textrm{est}}(t^\prime)} \textrm{d}t^\prime \label{eqrhoTWRpre1}\\
&\reviewermark{\approx c_0 \cdot \frac{ (\langle f_{M} \rangle + \delta f_{M}(t)) \cdot (\Delta t_\textrm{TpMr}+\Delta t_\textrm{MeTp})  }{2 \cdot \langle f_{M,\textrm{est}} \rangle} +  \textrm{const.} } \\ &= \reviewermark{\left( 1 + \frac{\langle f_{M} \rangle-\langle f_{M,\textrm{est}} \rangle}{\langle f_{M,\textrm{est}} \rangle} + \frac{\delta f_{M}(t)}{\langle f_{M,\textrm{est}} \rangle} \right)\cdot \frac{c_0 \cdot (2 \cdot \Delta t_\textrm{inst} + \mathcal{T}_\textrm{MeTp} + \mathcal{T}_\textrm{TpMr})}{2} +  \textrm{const.} } \\ &= \reviewermark{\left( 1 + \kappa + \delta \kappa(t) \right) \cdot \left( \rho_\textrm{inst} + \frac{  \mathcal{T}_\textrm{MeTp} + \mathcal{T}_\textrm{TpMr}}{2}  \right) +  \textrm{const.}}  \\
&= \rho_\textrm{inst}(t) + c_0 {\mathcal{T}}_\textrm{TWR}(t) \reviewermark{+ \kappa \cdot \rho_\textrm{inst} + \delta \kappa(t) \cdot \rho_\textrm{inst}(t)  + (\kappa + \delta \kappa(t)) \cdot c_0 {\mathcal{T}}_\textrm{TWR}(t)  +  \textrm{const.}}  \label{eqrhoTWR}  
	\end{align}
		The precise eq.~(\ref{eqrhoTWRpre1}) can be used to convert the phase observables to a non-instantaneous biased range $\rho_\textrm{TWR}$, even with a time-dependent frequency estimate $f_{M,\textrm{est}}(t)$. Under the assumption of a static estimate $\langle f_{M,\textrm{est}} \rangle$, and with eq.~(\ref{eqMasterPhaseM}), (\ref{eqMasterPhaseT}) and $ \Delta t_\textrm{TpMr,est} \approx \Delta t_\textrm{TpMr}$, the expression can be approximated as eq.~(\ref{eqrhoTWR}), which illustrates the coupling of frequency errors and the light-time correction effect. The first terms are the instantaneous range $\rho_\textrm{inst}$ and the light-time correction ${\mathcal{T}}_\textrm{TWR} = (\mathcal{T}_\textrm{MeTp} + \mathcal{T}_\textrm{TpMr})/2$, respectively. The third term describes a static scale factor error $\kappa = (\langle f_{M} \rangle-\langle f_{M,\textrm{est}} \rangle)/\langle f_{M,\textrm{est}} \rangle$ in the conversion from phase to range, while the term proportional to $\delta \kappa = \delta f_{M}(t)/\langle f_{M,\textrm{est}} \rangle$  accounts for laser phase variations, commonly known as laser frequency noise \cite{abich2019orbit}. The coupling of $\kappa$ or $\delta \kappa$ with the LTC in the fifth term is negligible compared to the same coupling with $\rho_\textrm{inst}$, because the magnitude of $c_0\mathcal{T}_\textrm{TWR}$ is below the millimeter level (cf.~table~\ref{tab:GReffects}).  The relevant aspect for the following sections is that the final Euclidean biased range can be computed as $\rho_\textrm{inst,TWR}=\rho_\textrm{TWR}-c_0 {\mathcal{T}}_\textrm{TWR}$.}

In order to compute the propagation time $\Delta t_\textrm{MeTp}$ from the master emission event (\textit{Me} on right plot of fig.~\ref{fig:Minkowski}) to the transponder reception (\textit{Tp} in fig.~\ref{fig:Minkowski}), the result of $\Delta t_\textrm{TpMr}$ is needed, as apparent from the following iterative equation
\begin{align}
\Delta t_\textrm{MeTp}^{(n+1)}(t_r) &= \frac{| \vec r_T(t_r-\Delta t_\textrm{TpMr}) - \vec r_M(t_r - \Delta t_\textrm{MeTp}^{(n)} - \Delta t_{TpMr}) |}{c_0} \nonumber \\
&\quad + \mathcal{T}_\textrm{GR}(\vec r_r = \vec r_T(t_r-\Delta t_\textrm{TpMr}), \vec r_e = \vec r_M(t_r - \Delta t_\textrm{MeTp}^{(n)} - \Delta t_\textrm{TpMr})) 
\label{eqFullLightConeEq4}
\end{align}
which we rigorously approximate\reviewermark{, with the same approach as utilized for eq.~(\ref{eqOWRanaPre}),} as
\begin{align}
\Delta t_\textrm{MeTp} &= \Delta t_\textrm{inst} + \mathcal{T}_\textrm{SR,MeTp} + \mathcal{T}_\textrm{GR,MeTp}\\
\mathcal{T}_\textrm{MeTp} 
&=  \frac{\Delta t_\textrm{inst} (\vec{d}_0\vdp\dot{\vec{r}}_T-2 \vec{d}_0\vdp\dot{\vec{r}}_M)+\Delta t_\textrm{inst}^2 \left(2 \vec{d}_0\vdp\ddot{\vec{r}}_M-\frac{\vec{d}_0\vdp\ddot{\vec{r}}_T}{2}\right)}{c_0}+\frac{\Delta t_\textrm{inst} \left(\left| \dot{\vec{r}}_T-2 \dot{\vec{r}}_M\right| ^2+(\vec{d}_0\vdp\dot{\vec{r}}_T)^2\right)}{2 c_0^2} \nonumber \\ 
&\quad+\frac{\Delta t_\textrm{inst}^2 \left(-2 \vec{d}_0\vdp\ddot{\vec{r}}_M (\vec{d}_0\vdp\dot{\vec{r}}_M-\vec{d}_0\vdp\dot{\vec{r}}_T)-4 \dot{\vec{r}}_M\vdp\ddot{\vec{r}}_M+2 \dot{\vec{r}}_T\vdp\ddot{\vec{r}}_M-\vec{d}_0\vdp\ddot{\vec{r}}_T \cdot \vec{d}_0\vdp\dot{\vec{r}}_T+\ddot{\vec{r}}_T\vdp\dot{\vec{r}}_M-\dot{\vec{r}}_T\vdp\ddot{\vec{r}}_T/2\right)}{c_0^2} \nonumber \\ 
&\quad+\frac{\Delta t_\textrm{inst} \left(\vec{d}_0\vdp\dot{\vec{r}}_T \left(2 \left| \dot{\vec{r}}_M\right| ^2+\left| \dot{\vec{r}}_T\right| ^2-2 \dot{\vec{r}}_T\vdp\dot{\vec{r}}_M\right)-2 \left| \dot{\vec{r}}_M\right| ^2 \vec{d}_0\vdp\dot{\vec{r}}_M\right)}{c_0^3} \\
&\quad + \frac{\mathcal{T}_\textrm{GR,TpMr} \cdot \vec d_0\vdp\dot{\vec{r}}_T - (\mathcal{T}_\textrm{GR,TpMr}+\mathcal{T}_\textrm{GR,MeTp}) \cdot \vec d_0\vdp\dot{\vec{r}}_M}{c_0}+\mathcal{O}(10^{-12}\,\textrm{m}/c_0).
\end{align}

The satellite state vectors, $\Delta t_\textrm{inst}$ and $\vec d_0 = (\vec r_M - \vec r_T)/|\vec r_M - \vec r_T|$ are evaluated at the reception time ($t_r$) and are the same as those needed to compute $ \mathcal{T}_\textrm{TpMr}$ with eq.~(\ref{eqOWRana}). The delay due to the atmosphere $\Delta t_\textrm{media}$ was omitted. The general relativistic contributions $\mathcal{T}_\textrm{GR} = \mathcal{T}_\textrm{PM} + \mathcal{T}_\textrm{HM} + \mathcal{T}_\textrm{SM}$ are evaluated at
\begin{align}
\mathcal{T}_\textrm{GR,TpMr} &= \mathcal{T}_\textrm{GR}(\vec r_r = \vec r_M(t_r),~\vec r_e = \vec r_T(t_r - \Delta t_\textrm{inst} - \Delta t_\textrm{inst} \vec d_0\vdp\dot{\vec{r}}_T/c_0) ) \\
\mathcal{T}_\textrm{GR,MeTp} &= \mathcal{T}_\textrm{GR}\left(\vec r_r = \vec r_T(t_r - \Delta t_\textrm{inst} - \Delta t_\textrm{inst} \vec d_0\vdp\dot{\vec{r}}_T/c_0 ),~\vec r_e = \vec r_M\left(t_r - \Delta t_\textrm{inst} \cdot \left( 2 c_0 + \vec d_0\vdp\dot{\vec{r}}_T - \vec d_0\vdp\dot{\vec{r}}_M \right)/c_0 \right) \right),
\end{align}
with the help of the Taylor expansion in eq.~(\ref{eqTaylorSC}).

It is noteworthy that the \reviewermark{leading term} in the TWR light-time correction
\reviewermark{
	\begin{align}
	c_0 \widehat{\mathcal{T}}_\textrm{TWR} & = c_0 \frac{\mathcal{T}_\textrm{MeTp} + \mathcal{T}_\textrm{TpMr}}{2}
	= -\frac{ | \vec{r}_B - \vec{r}_A| \cdot \dot{\rho}_\textrm{inst,OD}}{ c_0} + \textrm{const.} + \ldots
	\label{eqDominantTWR}
	\end{align}}
differs by a factor of two compared to the DOWR correction (cf.~eq.~(\ref{eqDOWRdominating})), whereby the static part has a similar magnitude (cf.~table~\ref{tab:GReffects}). 

\reviewersout{The last term, proportional to $\Delta f = f_M-f_T = 10\,\textrm{MHz}$, is usually negligible, since $\Delta f/f_M \approx 3.6 \cdot 10^{-8}$.}

\section{Requirements on light time correction precision}
\label{secRequirements}
It is sensible to require that the light time corrections $c_0 \mathcal{T}_\textrm{TWR}$ and $c_0 \mathcal{T}_\textrm{DOWR}$ are precise enough to not limit the precision of the instantaneous range, which is the measured biased range with subtracted light time correction. The precision of the instantaneous range $\rho_\textrm{inst}$ should ideally be limited by instrument noise and errors. 
Noise is driven by stochastic processes and can be described with spectral densities in the frequency domain. For instance, the noise requirement for the laser ranging instrument on GRACE FO is defined in terms of the amplitude spectral density (ASD), which is the square root of the power spectral density, as \cite{abich2019orbit}
\begin{align}
\textrm{ASD}[\rho_\textrm{LRI,req}] = 80\,\frac{\textrm{nm}}{\sqrt{\textrm{Hz}} } \sqrt{1+\left(\frac{3\,\textrm{mHz}}{f}\right)^2} \sqrt{1+\left(\frac{10\,\textrm{mHz}}{f}\right)^2}, \qquad 2\,\textrm{mHz}\leq f \leq 100\,\textrm{mHz}
\end{align}
while the corresponding requirement of the MWI reads \cite{kornfeld2019grace}
\begin{align}
\textrm{ASD}[\rho_\textrm{KBR}] = 2.62\,\frac{\mu \textrm{m}}{\sqrt{\textrm{Hz}} } \sqrt{1+\left(\frac{3\,\textrm{mHz}}{f}\right)^2}.
\end{align}
Deterministic or systematic errors manifest often as sinusoidal variations, so called tone errors. These should not exceed $\delta\rho = 1\,\mu \textrm{m}$ peak amplitude in GRACE FO measurements. This value is specified for the MWI at twice the orbital frequency ($f = 2 f_\textrm{orb} \approx 0.35\,\textrm{mHz}$) and for the LRI between $10 f_\textrm{orb} \leq f \leq 200 f_\textrm{orb}$ \cite{kornfeld2019grace}\footnote{\reviewermark{The $2f_\textrm{orb}$ KBR requirement is likely inherited and adopted from the GRACE mission \cite[p.~23]{grace1998grace}, while the higher LRI requirement band ($10 f_\textrm{orb} .. 200 f_\textrm{orb}$)  could be justified by the fact that other error sources like accelerometer or background model deficiencies limit the gravity field accuracy at lower frequencies. The authors recommend that both requirements are revised in future missions.}}. Although not strictly specified by the instruments, it is reasonable to require that the LTC has no sinusoidal errors above $1\,\mu \textrm{m}$ magnitude for all frequencies.

In the next sections, we illustrate the frequency content of time-domain signals with ASD plots, where the y-axis has units of $\rm{m} / \sqrt{\textrm{Hz}}$. These plots show the peak amplitude $\delta\rho$ of a sinusoidal variation with an amplitude of 
\begin{align}
\frac{\delta\rho}{\sqrt{2} \sqrt{\textrm{ENBW}}},
\end{align}
where ENBW is the equivalent noise bandwidth with units of Hertz. The ENBW depends on many parameters such as the length of the time-series, the sampling rate and the window function \cite{Heinzel.2002}. Since many gravity field recovery methods are using range rates, we recall that ASD values at a Fourier frequency $f$ with units of $\rm{m} / \sqrt{\textrm{Hz}}$ can be converted into the range rate domain with units $\rm{m} / (\textrm{s}\sqrt{\textrm{Hz}})$ by a multiplication with $2 \pi f$.

The actual in-orbit ASD of the LRI is well below the 80\,nm/$\sqrt{\textrm{Hz}}$ requirement as shown in \cite{abich2019orbit}, i.e. 
\begin{equation}
\textrm{ASD}[\rho_\textrm{LRI}]=\left\{
\begin{array}{@{}ll@{}}
15\,\textrm{nm}/\sqrt{\textrm{Hz}}, & f = 35\,\textrm{mHz} \\
0.3\,\textrm{nm}/\sqrt{\textrm{Hz}}, & f = 0.85\,\textrm{Hz}
\end{array}\right. 
\end{equation} 
which imposes stricter goals for the LTC precision at high frequencies.

\section{Validation of the analytical approximations for $\Delta t$}
\label{secValidationOWR}
In order to verify the equations for the light propagation time and our implementation of the software code, we performed a closed-loop \reviewermark{(i.e. backward-forward)} simulation using reduced-dynamic orbit data of  both GRACE Follow-On satellites in the Internetional Celestial Reference Frame (ICRF) from 5th February 2019 (GNI1B Release 04). One of the two satellites is designated as receiver with position $\vec r(t_r)$. At each epoch $t_r$ of the data, which has a sampling rate of 1\,Hz, the light propagation time $\Delta t = \Delta t_\textrm{inst} + \mathcal{T}_\textrm{SR} + \mathcal{T}_\textrm{GR}$  between the satellites is computed according to \reviewermark{eq.~(\ref{eqOWRanaPre})-(\ref{eqOWRanaGR}), which make use} of eq.~(\ref{eqTPM})-(\ref{eqTSR}) \reviewersout{and sec.~\ref{secLightTimeEquation}}.
With the propagation time $\Delta t$, we compute the photon emission position $\vec r_e$ and emission time $t_r - \Delta t$. Afterwards, we determine the vectorial coordinate speed of light $c_n \cdot \vec d_0$ pointing to the receiver (eq.~(\ref{eqCnFinal}) and (\ref{eqd0def})), which serves as the initial condition for a numerical integration of the equations of motion for photons (eq.~(\ref{eqGRequationmotion})) using the Adams-Bashforth-Moulton method \cite{shampine1997matlab}. The metric tensor used is based on a high-fidelity geopotential field, computed according to the models shown in table~\ref{tab:1}, and takes into account the vector potential due to Earth's spin. The integration is performed for a duration of $\Delta t$, which yields the photon path with an end position $\vec r_r^\prime$. If the analytical expressions to compute $\Delta t$ are correct, $\vec r_r^\prime$ and $\vec r_r$ should be identical. Hence, we define the error $\epsilon$ in the analytically-derived $\Delta t$ as
\begin{equation}
\epsilon = \left(\vec{r}_r^\prime-\vec{r}_r\right) \vdp \frac{\dot{\vec{r}}_{r}^\prime}{|\dot{\vec{r}}_{r}^\prime|} \approx \left(\vec{r}_r^\prime-\vec{r}_r\right) \vdp \vec{d}_0 \approx (\Delta t^\prime - \Delta t) \cdot c_0 
\end{equation}
which takes into consideration only the error in the propagation direction of the photon, since this contributes to the phase measurement in microwave or laser ranging. In other words, $\epsilon$ is the error of the computed $\Delta t$ with respect to the more accurate $\Delta t^\prime$.  

The lateral \reviewermark{part of the} displacement $\vec r_r^\prime-\vec r$ \reviewermark{is of} the order of 4\,$\mu$m \reviewermark{and} arises due to the light bending (cf.~sec.~\ref{sec:ltc_rel}), which has been omitted in our analytical \reviewersout{model} \reviewermark{approximation. By evaluating $\epsilon$, it can be shown that the bending - and omission of the bending in the analytical approximation - has a negligible effect on the phase measurement, since the longitudinal offset in propagation direction is very small and since the phasefront is, in good approximation, planar in the vicinity of $\vec r_r^\prime$}, i.e.,~the offset \reviewermark{$\vec r_r^\prime-\vec r$} vanishes when projected onto the propagation direction.

Due to the limited precision of double floating-point arithmetic, we perform the numerical integration in uniform co-moving coordinate frames, in order to have state vectors with small numerical values. This allows us to resolve even minor contributions within the light time correction.

\begin{figure}
	\centering
	\subfloat[][]{\includegraphics[scale=0.3]{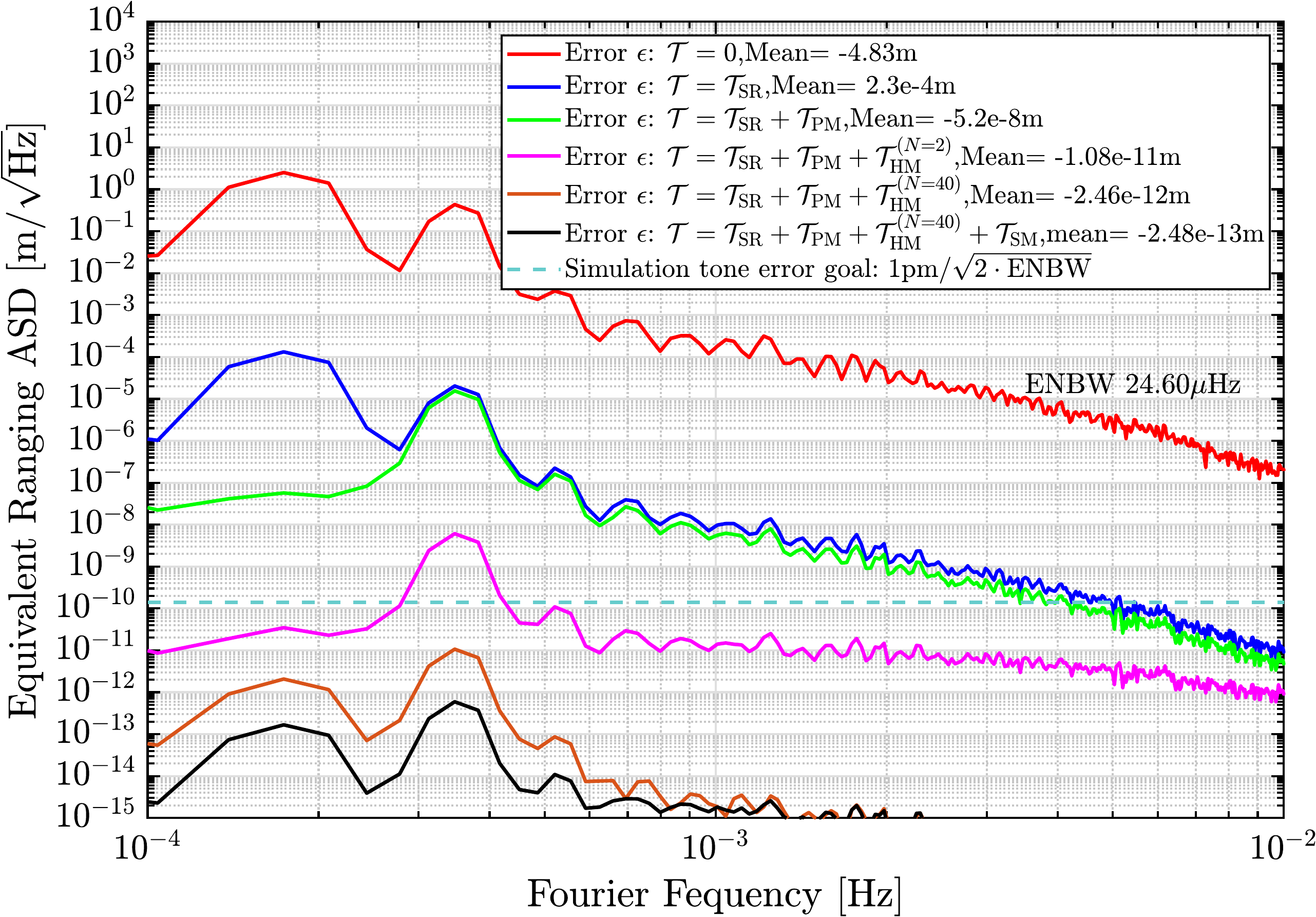}}\quad
	\subfloat[][]{\includegraphics[scale=0.3]{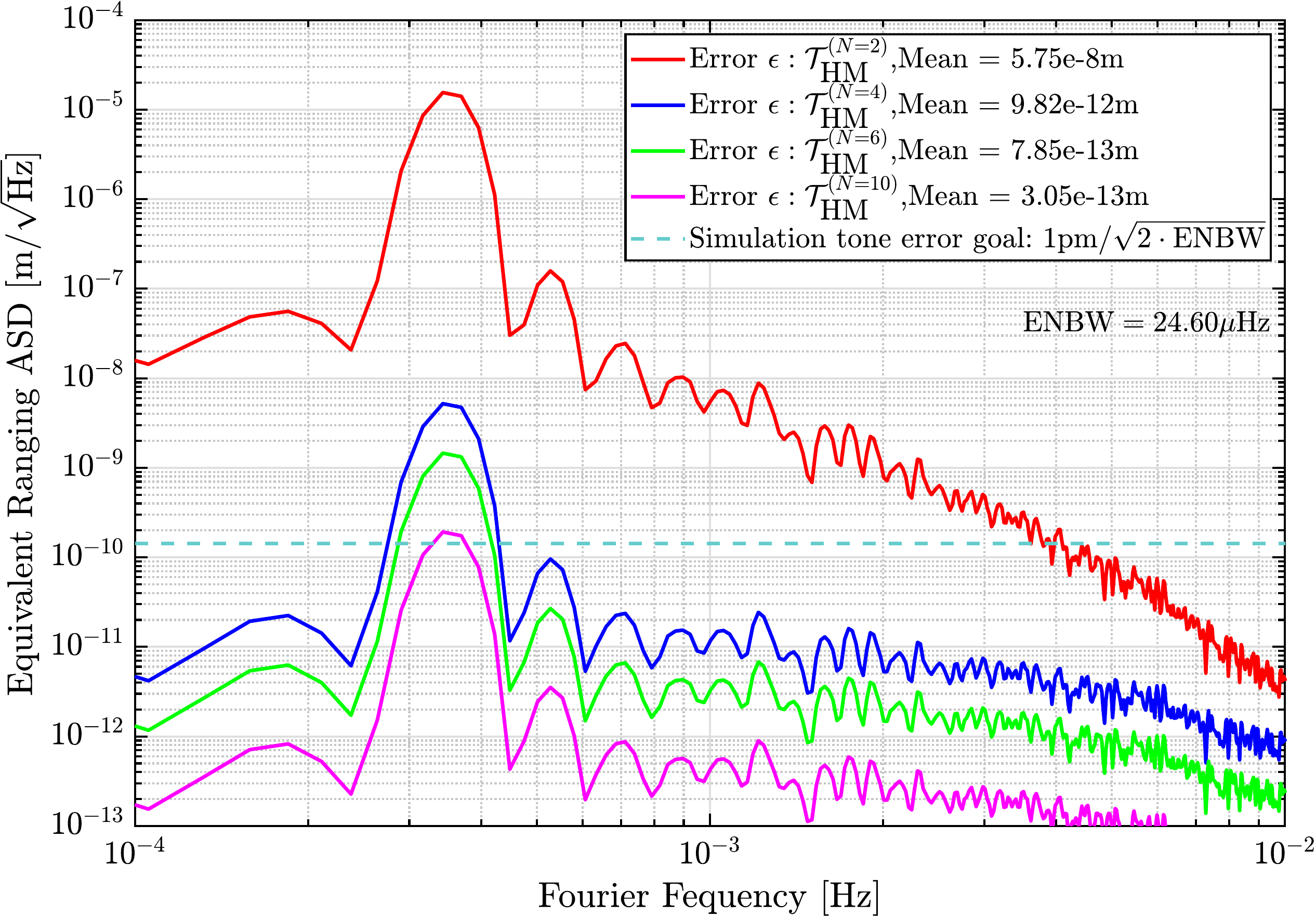}}\\
	\subfloat[][]{\includegraphics[scale=0.3]{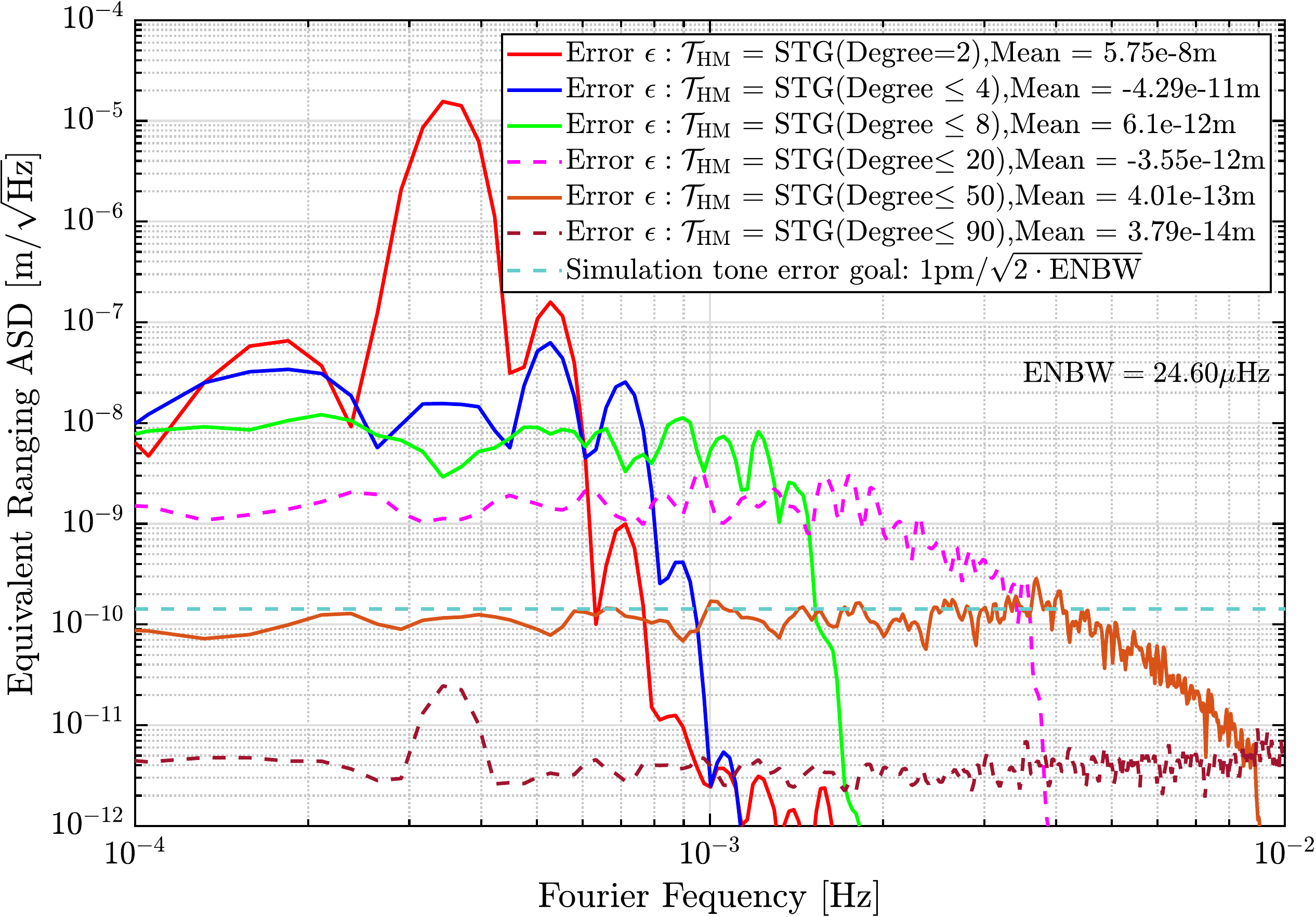}}\quad
	\subfloat[][]{\includegraphics[scale=0.3]{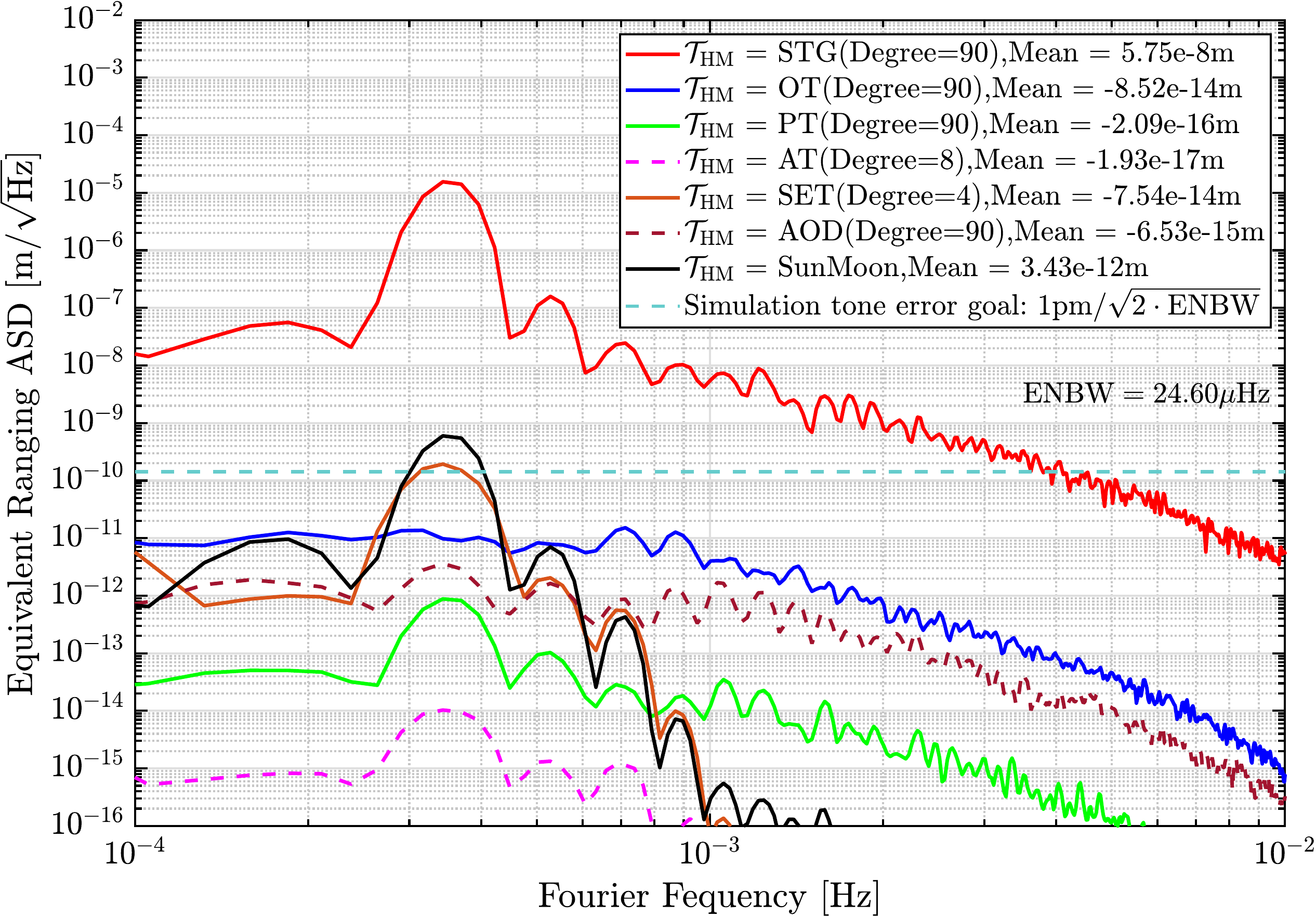}}

	\caption{ Amplitude spectral density plots of the model error $\epsilon$ and of the term $\mathcal{T}_\textrm{HM}$. \textbf{a)} The first six traces show the model error $\epsilon$ for different contributors in the light time correction $\mathcal{T}$. The model error $\epsilon$ as a function of the number of sampling points $N$ of the path integral (eq.~(\ref{eqTHM})) is shown in subfigure \textbf{b)}, while the influence of the truncation degree for the SH expansion of the gravitational potential is illustrated in \textbf{c)}. Subfigure \textbf{d)} shows the ASD of a time series of $\mathcal{T}_\textrm{HM}$, where only a single gravitational potential model from table~\ref{tab:1} was used. All subfigures use the Nuttall4a window function. The equivalent peak height of a sinusoidal variation with 1 picometer amplitude is visualized as green dashed line in all four plots.}
	\label{fig_validation}
\end{figure}

The result of the one-way ranging validation, i.e.~the time series of $\epsilon$, is shown in the spectral domain in figure~\ref{fig_validation}a). The upper-most  trace in red shows the error $\epsilon$, if special and general relativistic effects are omitted in the calculation of the light travel time $\Delta t$, which means $\Delta t = t_\textrm{inst}$. Considering $\mathcal{T}_\textrm{SR}$ yields the blue trace. The general relativistic contribution to the light propagation shows two sinusoidal variations at once and twice the orbital frequency and a continuous spectral content decaying towards higher frequencies. The peak at \reviewermark{t}he orbital frequency is caused by the radially symmetric gravity field ($\mathcal{T}_\textrm{PM}$), while the higher moments cause the twice per revolution peak and the continuous part.

Since the spectral plots conceal the DC component, the mean value of $\epsilon$ is provided in the legend. The figure confirms that the different contributions in the propagation time indeed reduce the error $\epsilon$ down to a mean level of $2.5 \cdot 10^{-13}\textrm{m}/c_0$, with fluctuations well below $1\,\textrm{pm}/\sqrt{\textrm{Hz}}/c_0$. The remaining peaks apparent at once and twice the orbital frequency from sinusoidal variations (tones) are not described properly with units of a spectral density plot (cf.~sec.~\ref{secRequirements}). These variations have a peak magnitude in the time-domain of less than 1\,picometer (green dashed line in fig.~\ref{fig_validation}a), if $\mathcal{T}_\textrm{PM}$ and $\mathcal{T}_\textrm{HM}$ are considered . 

The contribution of the general relativistic correction $\mathcal{T}_\textrm{SM}$ due to Earth's spin moment is present predominantly at once and twice the orbital frequency, but with a negligible magnitude (difference between brown and black trace). Hence, $\mathcal{T}_\textrm{SM}$ can be safely omitted from now on.

The dependence of the model error $\epsilon$ on the sampling point number $N$ in eq.~(\ref{eqTHM}) is shown in fig.~\ref{fig_validation}b), while fig.~\ref{fig_validation}c) visualizes the effect of the truncation degree for the SH expansion of the gravitational potential. The actual signal $\mathcal{T}_\textrm{HM}$ for different individual models of the gravitational potential (cf.~table~\ref{tab:1}) is depicted in \ref{fig_validation}d). In general, fig.~\ref{fig_validation} can be used to decide which models and parameters are required for a particular accuracy level in the computation of the light time correction.

Although this section showed only one-way ranging results, most of the findings are also applicable for the TWR and DOWR combinations, since these are formed by the average of two one-way ranging results. Only $\mathcal{T}_\textrm{SM}$ and some terms in $\mathcal{T}_\textrm{SR}$ flip signs between the two opposite directions, which means they are canceling to a large extent in the TWR and DOWR case.

A result of this analysis is that the following parameters of $\mathcal{T}_\textrm{HM}$ are sufficient to meet  the precision requirements formulated in sec.~\ref{sec:Comparison} and sec.~\ref{sec:Enhancing}: SH degree of the static gravity should be $\geq50$, while a Solid Earth Tide (SET) model with degree 4 is sufficient;  the path integral should be approximated with N$\geq10$ and direct tidal accelerations should be taken into account at least from Sun and Moon.

\section{Comparison with GRACE and GRACE FO Light Time Correction}
\label{sec:Comparison}
We compared the method to derive the light time correction presented herein with the light time correction values in the level-1b data of the GRACE and GRACE Follow-On mission\reviewermark{s}. These values are provided in the KBR1B and LRI1B datasets alongside with the actual biased range. The most recent version of the GRACE data is release 03 (RL03), which is available only for the SCA1B and KBR1B data products, while for all other products RL02 is the most recent version \cite{PODAAC1}. Details on the processing of GRACE data can be found in \cite{wu2006algorithm}. The GRACE Follow-On data is available in version RL04 by the time of the writing \cite{GFOdataRL04}. 

For the GRACE data, the GNV1B orbit data is rotated from the terrestrial to the celestial frame by a rotational matrix formed according to the IAU-2000 standard using Earth orientation parameteres \cite{petit2010iers}. The sampling rate of the orbit data is 0.2\,Hz, hence, it is directly compatible with the KBR1B data. Since the LTC for microwave ranging needs to be referred to the antenna phase center (APC), the position of the phase center in the satellite frame, as provided by VKB1B\footnote{value from the year~2012 in the sequence of events file}, is rotated using the star camera SCA1B product into the ICRF. The COM-APC offset in the ICRF is added onto the rotated GNV1B data in order to obtain the position and velocity of the APC on each SC in the ICRF. The acceleration vector of the APC is approximated by the center-of-mass acceleration from force models, which is justified, since the angular motion of the APC on time scales of the light propagation time is negligible. The APC state vectors are used to derive the one-way LTCs $\mathcal{T}_{AeBr}$ and $\mathcal{T}_{BeAr}$ (eq.~(\ref{eqOWRana})), which are further combined using eq.~(\ref{eqTDOWRcombin}) into $\mathcal{T}_\textrm{DOWR}$ with K- and Ka-band frequencies from the USO1B dataset. 

\begin{figure}
	\centering
     \includegraphics[scale=0.3]{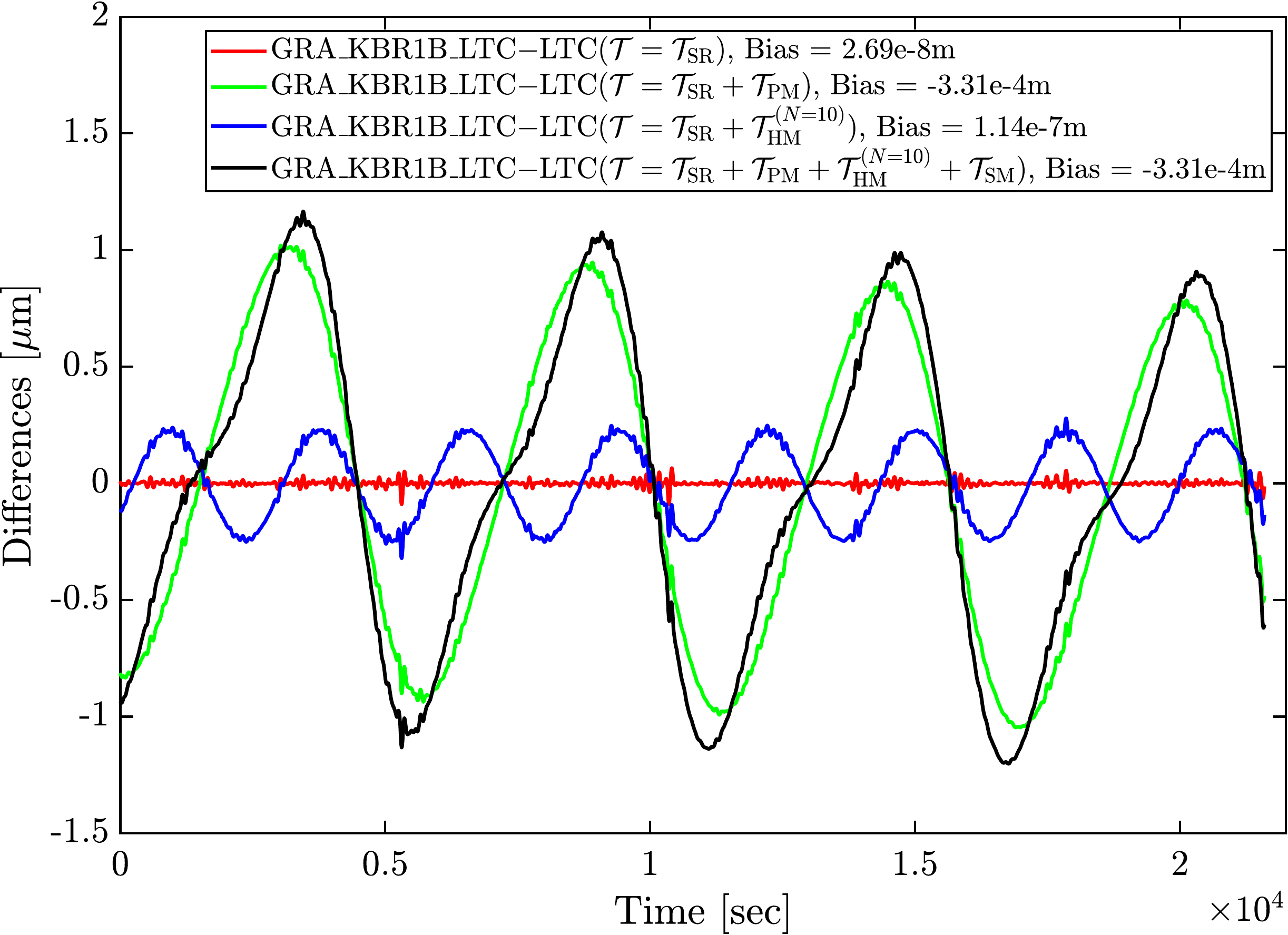} \quad 
	\includegraphics[scale=0.3]{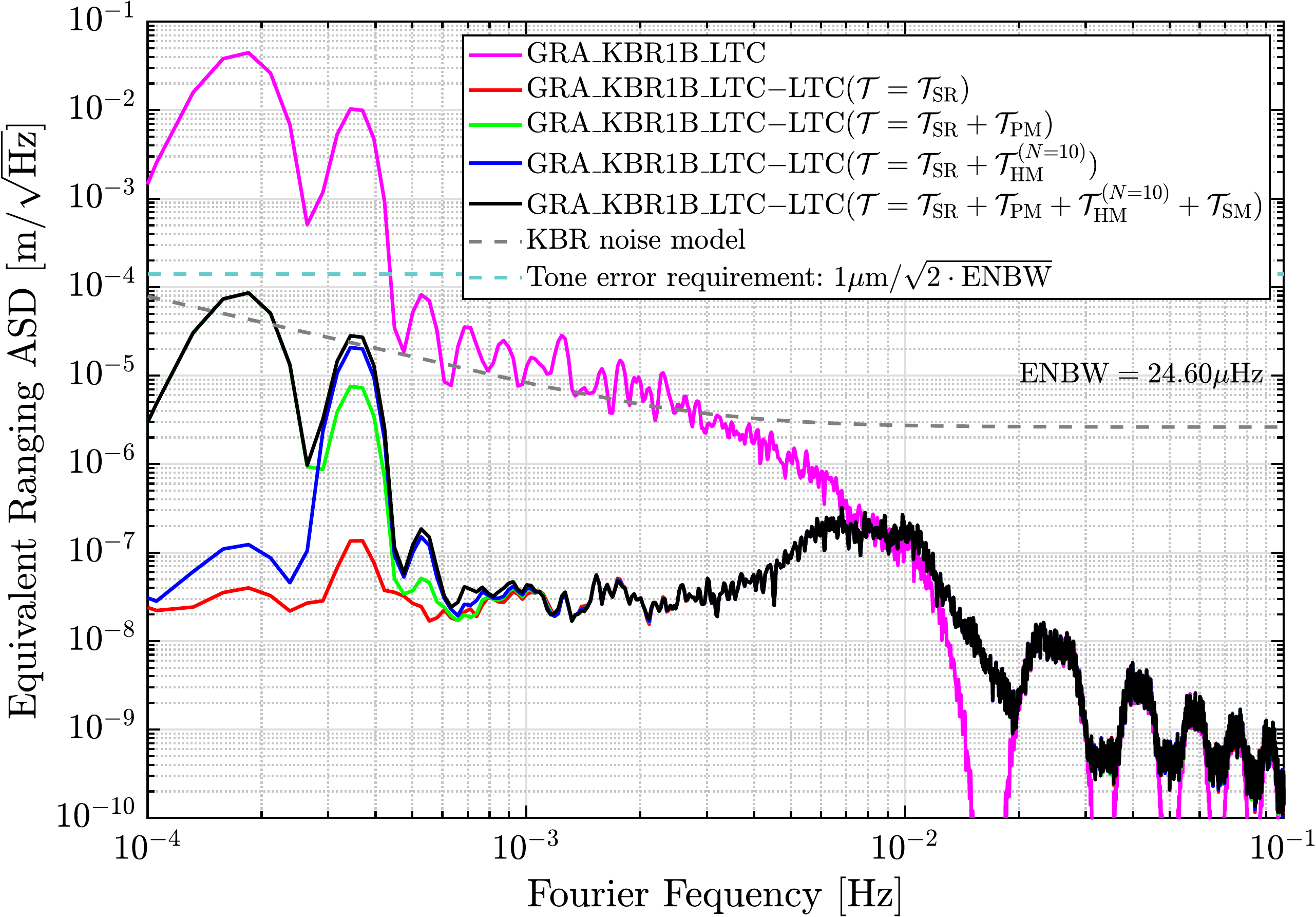}
	\caption{Comparison between GRACE level-1b light time correction and $\mathcal{T}_\textrm{DOWR}$ (eq.~(\ref{eqTDOWRcombin})) using different degrees of accuracy in the time (left) and spectral (right) domain. The traces on the left plot have been centered around zero by subtracting a bias shown in the legend. The difference is minimal when only the special relativistic effect is considered in $\mathcal{T}_\textrm{DOWR}$. The dominating amplitudes and the mean values are provided table~\ref{tab:2}. }
	\label{fig:ComparisonGRACE}
\end{figure}

The difference between the light time correction from GRACE level-1b KBR data (GRA\_KBR1B\_LTC) and $c_0 \cdot \mathcal{T}_\textrm{DOWR}$ (eq.~(\ref{eqTDOWRcombin})) with four different degrees of accuracy is shown in fig.~\ref{fig:ComparisonGRACE}. The data used spans the GPS time between 00:00 and 06:00 on December 1st, 2008.  Since the differences are minimal when only the special relativistic correction $\mathcal{T}_\textrm{SR}$ is used (red trace), it is reasonable to assume that general relativistic contributions were omitted in the GRACE level-1b light time correction. The omission error is dominated by the sinusoidal variation at the orbital frequency, however, with an amplitude of approx.~1\,micrometer, i.e.~close to the tone error requirement discussed in sec.~\ref{secRequirements} for GRACE Follow-On.

The GRACE level-1b LTC shows some artifacts above 10\,mHz (magenta trace on the right subplot in fig.~\ref{fig:ComparisonGRACE}). However, these are well below the KBR noise level and should not impede the gravity field recovery.

\begin{figure}
	\centering
	\includegraphics[scale=0.3]{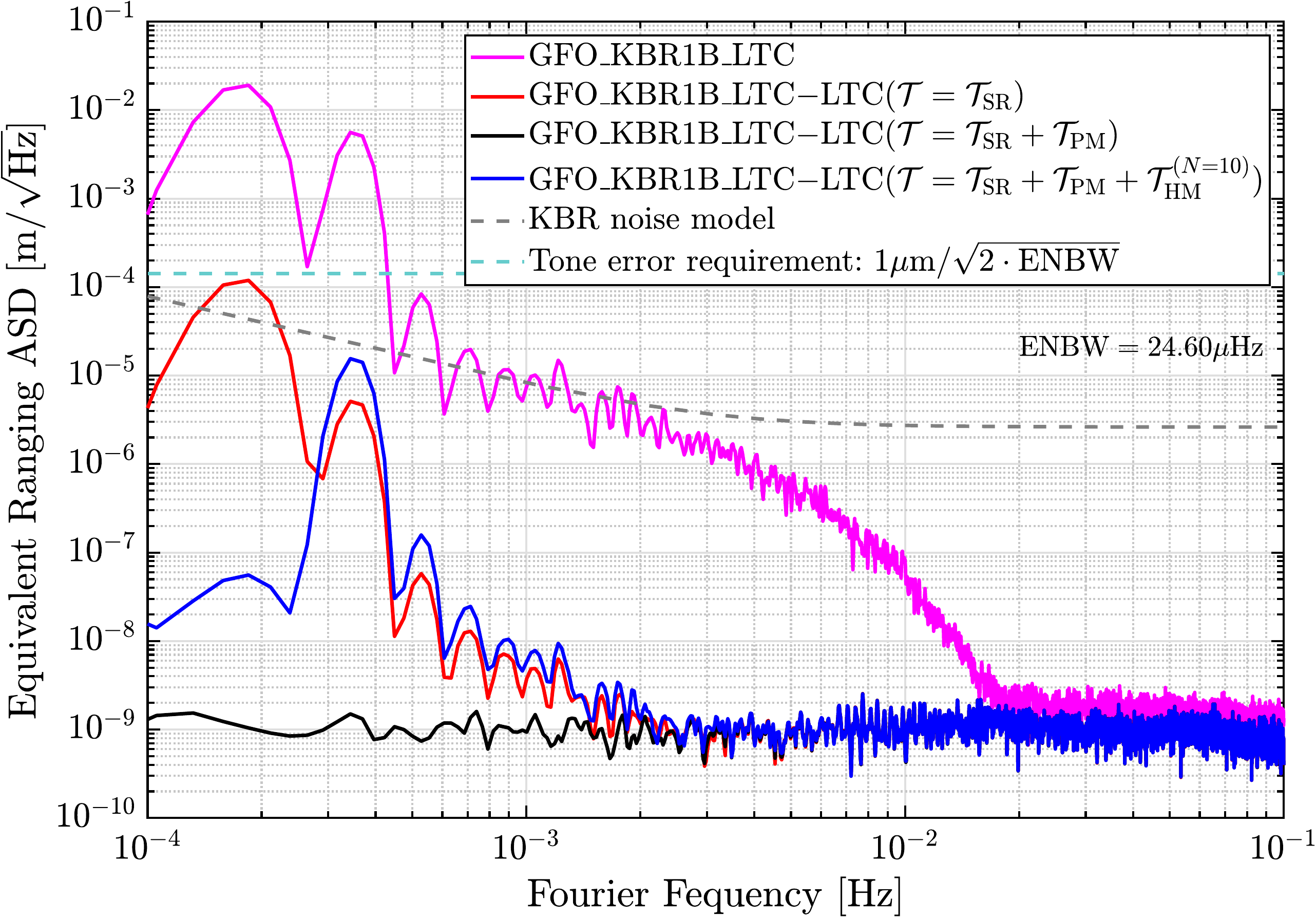}
	\includegraphics[scale=0.3]{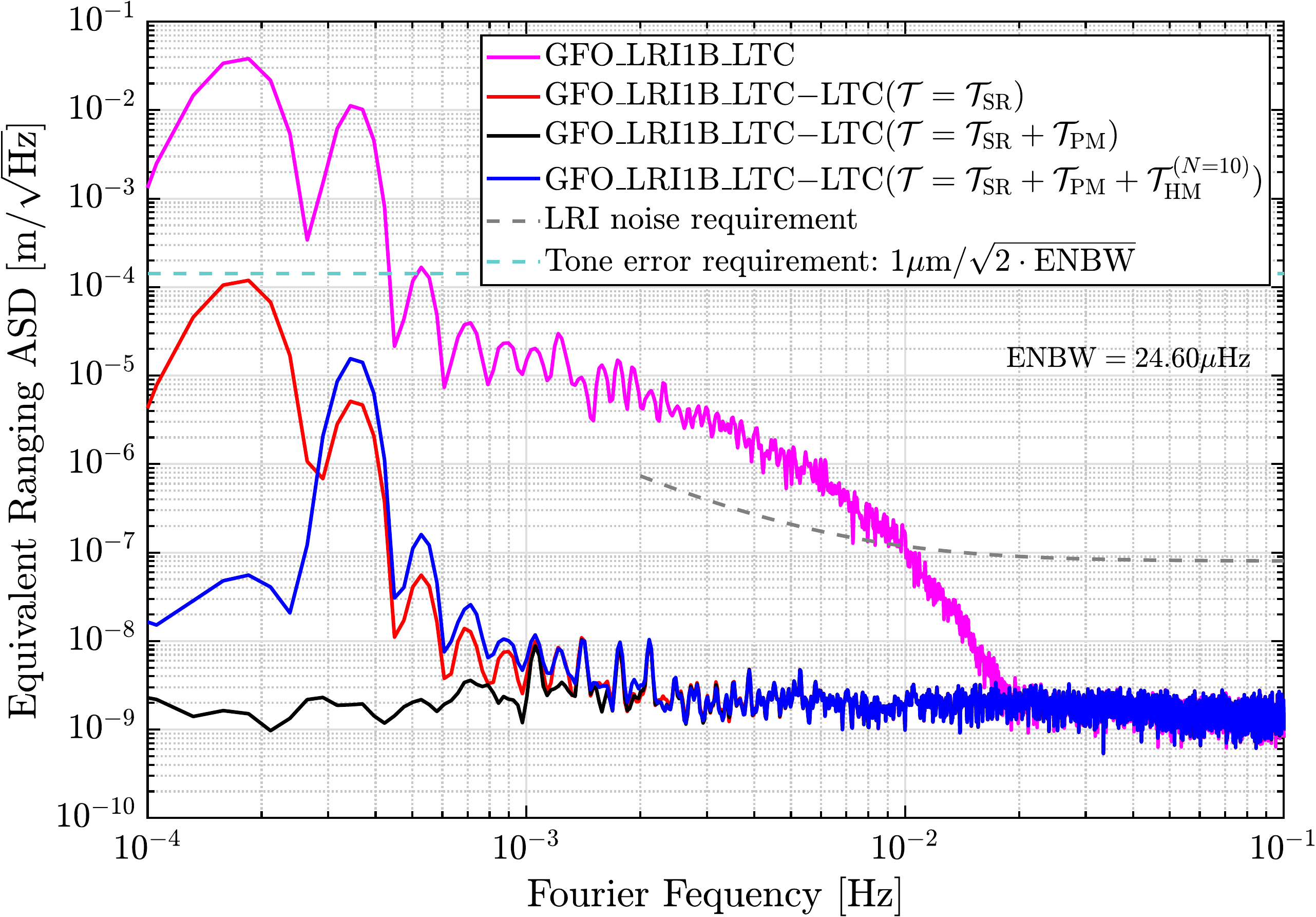}
	\caption{Comparison between GRACE-Follow-On(GFO) level-1b KBR/LRI light time correction and $\mathcal{T}_\textrm{DOWR}$ (left) and $\mathcal{T}_\textrm{TWR}$ (right) using different degrees of accuracy. The dominating amplitudes and the mean values for the different traces are provided table~\ref{tab:3}. }
	\label{fig:ComparisonGRACEFO}
\end{figure}

For GRACE Follow-On, an additional orbit data product called GNI1B is available, which provides the satellite state in the ICRF and can be used instead of the transformed GNV1B data. The sampling rate is 1\,Hz, which means that results need to be downsampled to the KBR and LRI rates of 0.2 and 0.5\,Hz, respectively. A comparison with different degrees of accuracy for the light time correction is shown in fig.~\ref{fig:ComparisonGRACEFO} for February 5th, 2019. It is evident that the LTC in GRACE FO takes into account the general relativistic effect $\mathcal{T}_\textrm{PM}$ due to the central field (degree 0), but not the higher moments $\mathcal{T}_\textrm{HM}$. The omission error is present predominantly at twice the orbital frequency with a peak amplitude of approx.~0.1\,$\mu$m (blue trace), thus well below the discussed requirement from sec.~\ref{secRequirements}. The differences between $c_0 \cdot \mathcal{T}_\textrm{TWR}$ and the RL04 LTC in fig.~\ref{fig:ComparisonGRACEFO} are limited to a level of a few nm/$\sqrt{\textrm{Hz}}$, which is well below the LRI noise requirement.

However, the actual LRI in-orbit noise is close to 1\,nm/$\sqrt{\textrm{Hz}}$ at Fourier frequencies around 0.1\,Hz, hence, we study the limits of the LTC precision and propose potential improvements for the LTC in the next section.

\section{Enhancing the Light Time Correction Accuracy}
\label{sec:Enhancing}
In order to understand the current limit of the LTC precision of a few nm/$\sqrt{\textrm{Hz}}$, we reproduced the light time corrections provided in the GRACE Follow-On RL04 data. In a first step (step~1), the classical light time equation was solved iteratively to obtain the absolute light travel time $\Delta t$, and, in a second step (step~2), the instantaneous contribution ($\Delta t_\textrm{inst} = |\vec r_A - \vec r_B|/c_0$) was removed from $\Delta t$ in order to obtain the one-way corrections $\mathcal{T}$, which are further combined into $\mathcal{T}_\textrm{DOWR}$ or $\mathcal{T}_\textrm{TWR}$.

\begin{figure}
	\centering
	\includegraphics[scale=0.3]{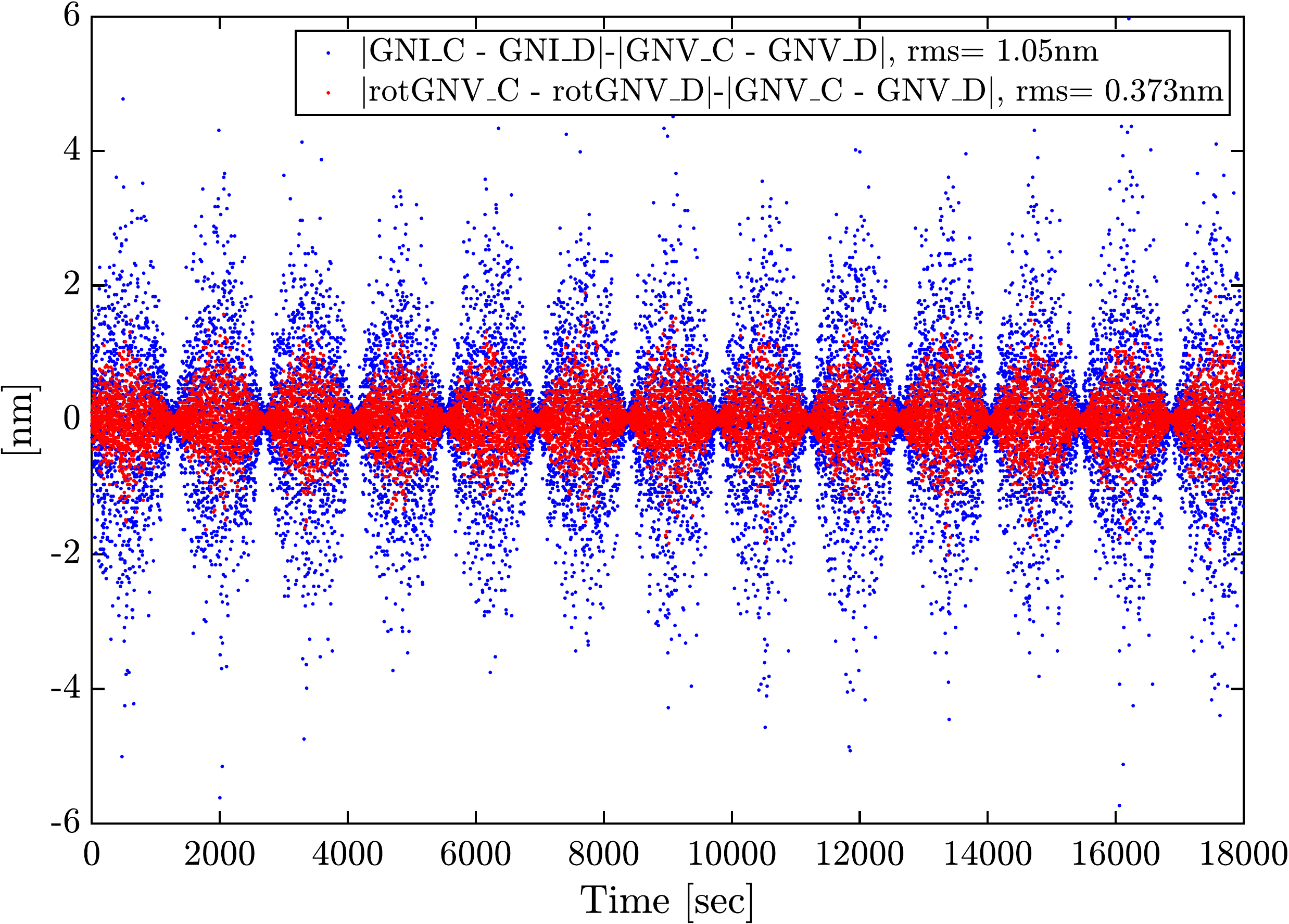}
	\includegraphics[scale=0.3]{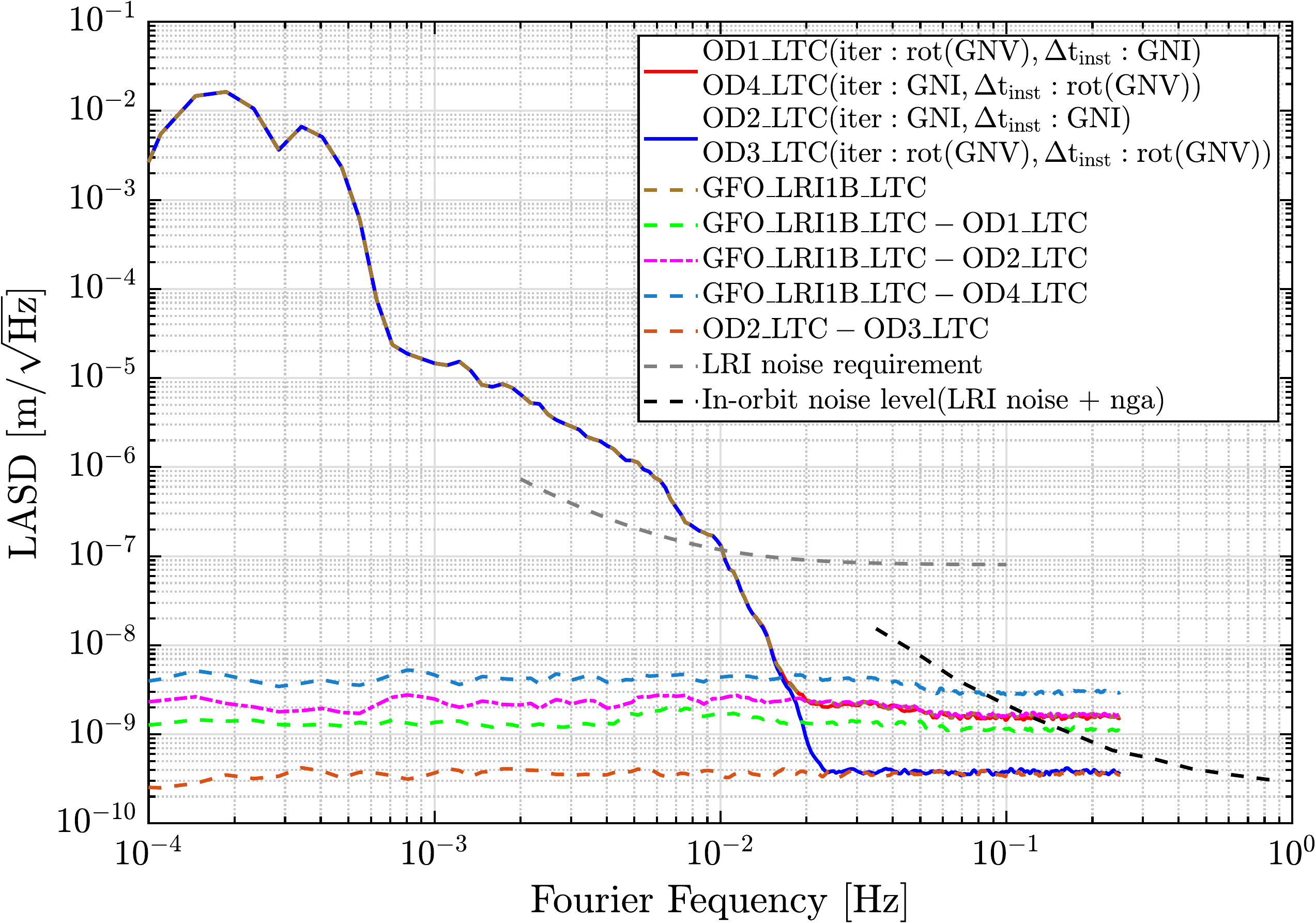}
	\caption{(Left plot:) Difference in inter-satellite distance between satellite C and D for different orbit data products. GNI and GNV are the RL04 datasets, while rotGNV denotes a dataset, which has been rotated by the authors from ITRF to ICRF. (Right plot:)  Spectral density of the LTC signal (red and blue traces) and LTC differences for different orbit data sets. The LTCs have been computed in four different cases (OD1..OD4), which are based in different orbit data sets for step~1 (iter: solving for $\Delta t$ iteratively) and step~2 (calculating $\Delta t_\textrm{inst}$ and computing $\mathcal{T} = \Delta t - \Delta t_\textrm{inst}$). This plot was created with a log-scale amplitude spectral density (LASD) method, which produces smooth traces also at high frequencies \cite{trobs2006improved}.}
	\label{fig:OrbitProducts}
\end{figure}

We noted a slight inconsistency in the instantaneous \reviewermark{Euclidean} inter-satellite distance between GNI1B or GNV1B products, which shows rms differences three times higher compared to our method to rotate the GNV1B data into the ICRF (cf.~left panel in fig.~\ref{fig:OrbitProducts}). The precision limit of our method is the resolution of the double floating-point arithmetic, i.e.~the computation error of the product of rotation matrix and position vector.

We could reproduce the light time correction of RL04 data with smallest deviations, if we used different orbit sets in step~1 and for the calculation of $\Delta t_\textrm{inst}$ in step~2 (cf.~green dashed trace on right subplot of fig.~\ref{fig:OrbitProducts}). However, using consistent orbit sets for both steps results in a slightly lower noise for the light time correction (solid blue trace). The consistent data sets could be GNI for both steps (denoted as orbit data OD2 in the plot), or the rotated GNV data (denoted as orbit data OD3). A difference between both cases is not apparent in the spectrum, hence, the plot shows a single solid blue trace for both cases. The dashed black trace on the right plot of fig.~\ref{fig:OrbitProducts} depicts the actual in-orbit measurements of the LRI \cite{abich2019orbit}, which contains the instrument noise but also some variations due to non-gravitational accelerations (nga) for the shown frequencies \cite{misfeldt2019}.

\begin{figure}
	\centering
	\includegraphics[scale=0.3]{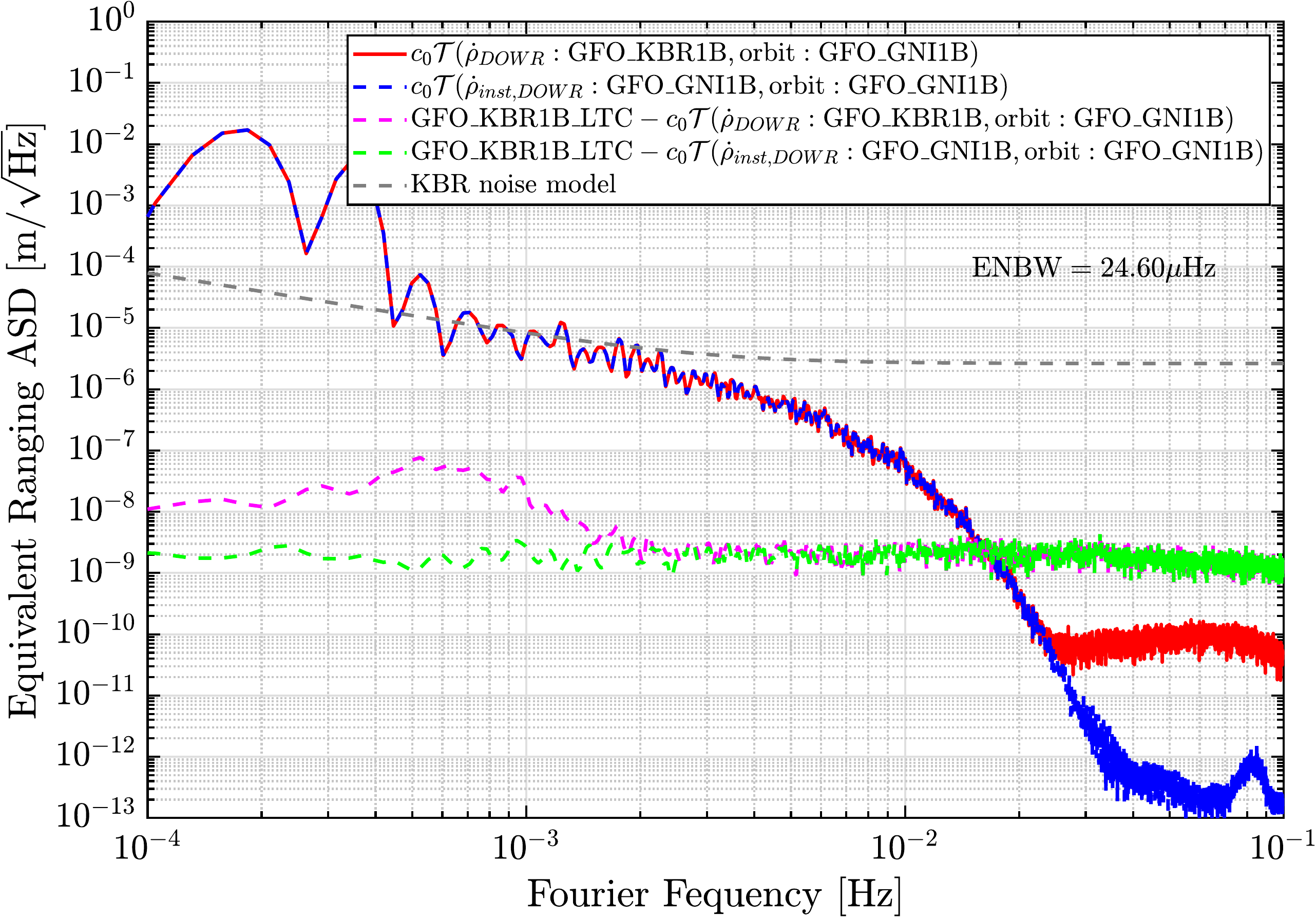} 
	\includegraphics[scale=0.3]{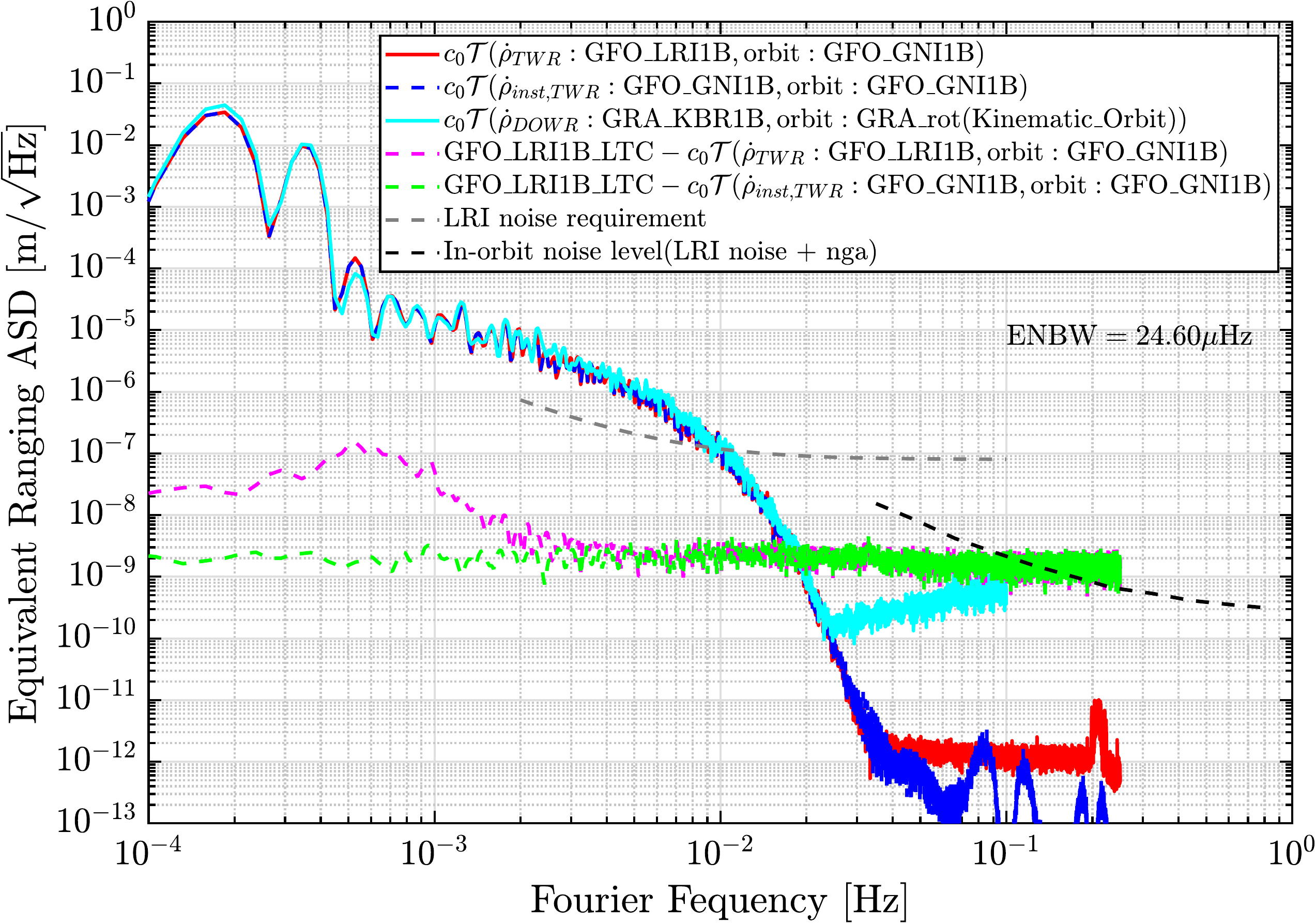}
	\caption{Amplitude spectral density (ASD) of the differences of the LTC with different inter-satellite range rate data for one day in August 1 2019 with logarithmic scaled frequency axis.   \reviewersout{ Nuttall4a is used as the window function for spectral estimation (ENBW = 24.60$\mu$Hz)} (\reviewermark{Left:}) comparison for the LTC range of KBR, (\reviewermark{Right:}) comparison for the LTC range of LRI. In addition, the cyan blue trace in the right subplot shows the LTC from a kinematic orbit of GRACE (December 1, 2008).}
	\label{fig:EnhancedLTC}
\end{figure}

The LTC accuracy can be improved further -\,well below the sensitivity of the LRI\,- by using the analytical expressions for $\mathcal{T}$ as discussed in sec.~\ref{secLTCinDOWR} and \ref{secLTCinTWR}, where the dominating terms in the single-path are proportional to $\vec d_0 \vdp \dot{\vec r}_{A/B}$, or in the final DOWR and TWR combination $\mathcal{T}_\textrm{DOWR/TWR} \propto \vec d_0 \vdp (\dot{\vec r}_{A} - \dot{\vec r}_{B}) \propto  \dot{\rho}_\textrm{inst,OD} $ (cf.~eq.~(\ref{eqDOWRdominating}) and (\ref{eqDominantTWR})). If the satellite velocity vectors $\dot{\vec{r}}_{A/B}$ are derived as the time-derivative of the satellite position state vector, the accuracy of the LTC is limited to the nm/$\sqrt{\textrm{Hz}}$ level. However, if the velocity state vectors of GNI1B are used, the LTC noise is highly reduced as shown by the blue traces in left and right plot of fig.~\ref{fig:EnhancedLTC}. This results from the fact that GNI and GNV data is based on reduced-dynamic orbit determination\reviewermark{, where the variational equations include the velocity state \cite{bertiger2010single, BERTIGER2020469}}\reviewersout{, and purely kinematic solutions show a higher noise, as shown by the magenta trace for GRACE, where kinematic orbit data is publicly available}.

It is noteworthy that the instantaneous range rate $\dot{\rho}_\textrm{inst,OD}$, which appears in the first order approximation of the LTC (eq.~(\ref{eqDOWRdominating}) and (\ref{eqDominantTWR})), dominates the noise in the LTC. Fortunately, the instantaneous range rate is approximately the same as the more precise measured range rate from LRI or KBR with only a minor light time correction from an orbit product, i.e.
\reviewermark{
	\begin{align}
	\dot{\rho}_\textrm{inst,TWR} \approx \dot{\rho}_\textrm{TWR}- \frac{\textrm{d}}{\textrm{d} t} \frac{|\vec{r}_A - \vec{r}_B| \cdot \dot{\rho}_\textrm{inst,OD}}{c_0}, \qquad \dot{\rho}_\textrm{inst,DOWR} \approx \dot{\rho}_\textrm{DOWR}- \frac{\textrm{d}}{\textrm{d} t} \frac{|\vec{r}_A - \vec{r}_B| \cdot \dot{\rho}_\textrm{inst,OD}}{2 c_0}. \label{eqLTCusingRanging}
	\end{align}
}
Thus, if $\dot{\rho}_\textrm{inst,OD}$ from the orbit product is replaced with $\dot{\rho}_\textrm{inst,TWR}$ or $\dot{\rho}_\textrm{inst,DOWR}$ in the dominating term of the LTC, the resulting LTC becomes almost independent of the orbit product. The result exhibits very low noise at high-frequencies (red trace on the right plot in fig.~\ref{fig:EnhancedLTC}) that is comparable to the pure GNI LTC (dashed blue trace). The deviations below 2\,mHz are caused by differences between ranging and orbit data, and it is reasonable to assume that the results using eq.~(\ref{eqLTCusingRanging}) are more accurate than the LTC based purely on orbit data. 

\reviewermark{
	Moreover, the above replacement allows us to use even kinematic orbit products for the LTC calculation with acceptable high frequency noise (cyan blue trace in fig.~\ref{fig:EnhancedLTC} for GRACE data \cite{zehentner2013kinematic}). For that trace, the high frequency noise above 25\,mHz is driven by the KBR ranging noise. Kinematic orbits \reviewersout{in general} are sometimes regarded as more appropriate for gravity field recovery \cite{naeimi2017global}, since they do not rely on a-priori gravity field information. 
}

Finally, we note that the most accurate way to determine the instantaneous biased range $\rho_\textrm{inst}$ is to update the LTC in the process of combined orbit determination and gravity field recovery with the most current orbit estimate in each iteration. In other words, one can consider to use the non-instantaneous biased range as observation and shift the conversion by means of the LTC into the process of precise orbit determination and gravity field recovery, where the LTC is updated iteratively.

\section{Summary \& Conclusions}

The Laser Ranging Interferometer aboard GRACE Follow-On demonstrated for the first time laser ranging between satellites in a gravimetric satellite mission. This enables inter-satellite biased range observations with an unprecedented noise level of 1\,nm/$\sqrt{\textrm{Hz}}$ at the highest frequency in the level-1b data (0.25\,Hz), or even 0.2\,nm/$\sqrt{\textrm{Hz}}$ at the highest frequency of the level-1a data (5~\,Hz). 

The biased range observation needs to be corrected for the effect of the finite speed of light in order to obtain the instantaneous range between the spacecraft, which is the quantity utilized in the gravity field recovery process. It is natural to seek methods to compute the light-time corrections with a higher precision in order to not limit the observations of the GRACE Follow-On LRI, and potentially also of future instruments and missions.

In this paper, we revisited the calculation of the light time correction from first principles within the Post-Newtonian approximation of general relativity, taking into account state-of-the-art geopotential models. We have separated the total light time correction $\mathcal{T}$ into the contribution from special relativity $\mathcal{T}_\textrm{SR}$ and the general relativistic component into the effect from the scalar central field of the Earth ($\mathcal{T}_\textrm{PM}$, SH degree 0), from higher moments of the gravity potential, which includes direct tidal accelerations, $\mathcal{T}_\textrm{HM}$, and from the much smaller vector potential due to Earth's spin moment $\mathcal{T}_\textrm{SM}$. The analytical formulas were verified against the light travel time obtained by numerically integrating the equations of motion of photons.

We studied in sec.~\ref{secValidationOWR} the influence of different geopotential models onto $\mathcal{T}_\textrm{HM}$, showing that to reach tone-errors below 1\,pm amplitude in the LTC, one should consider the effect from the Sun and the Moon, as well as from Solid Earth tides. In order to achieve a noise level in the light time correction below 100\,pm/$\sqrt{\textrm{Hz}}$, the SH degree of the static gravity field should be above 50 and the light path between satellites needs to be sampled with more than 10 points.

We showed that the GRACE light-time correction in RL02 does not consider general relativistic effects, while GRACE Follow-On RL04 data takes into account general relativity with a radial-symmetric field ($\mathcal{T}_\textrm{PM}$). The omission of $\mathcal{T}_\textrm{HM}$ causes predominantly a sinusoidal error with a peak amplitude well below 1\,$\mu$m at twice the orbital frequency. The LTC in the official RL04 data is limited to a noise level of a few nm/$\sqrt{\textrm{Hz}}$ arising from numerical floating point precision and due to the fact that two slightly inconsistent orbit products (GNI and GNV) are used in each step. This level of LTC precision is comparable to the LRI instrument noise at the highest frequencies in the level-1b data.

The numerical accuracy can be easily improved to 1\,nm/$\sqrt{\textrm{Hz}}$ at high frequencies by using the same orbit product in both steps. However, we recommend to use the here proposed analytical formulas as these are numerically a few orders of magnitude more accurate, as the absolute LTC accuracy depends on the models and on the orbit product quality. 

In the end it was pointed out that, if the analytical formulas are employed, the dominating term of $\mathcal{T}_\textrm{TWR}$ or $\mathcal{T}_\textrm{DOWR}$ can be rewritten in terms of the measured range rate from LRI or KBR, which means the LTC becomes to first order independent of the orbit product. Hence, kinematic orbit products that suffer higher noise can be used to compute the LTC as well.

The here presented methods to calculate the light time correction for microwave and laser ranging can be readily applied to simulated and available flight data. The analytical approximations were truncated at picometer level, which is well below the requirements for the current GRACE Follow-On mission, but may be of interest in studies for future missions.
\\~\\
{\footnotesize
	\noindent
	\textbf{Acknowledgements}
	Yihao Yan acknowledges the China Scholarship Council for scholarship support and expresses his gratitude to the GRACE-Follow-On LRI group members at the Albert-Einstein-Institute in Hannover, Germany. The authors thank Changqing Wang for the support of validating some force models. This work was supported by the National Natural Science Foundation of China (projects no. 41704013). This work was supported by the Max-Planck-Society and the Chinese Academy of Sciences within the LEGACY (``Low-Frequency Gravitational Wave Astronomy in Space'') collaboration (M.IF.A.QOP18098).}
\\~\\
{\footnotesize
	\noindent
	\textbf{Author Contributions}
	Vitali M\"uller conceived the research idea and performed preliminary computations. Yihao Yan refined the models, developed the computational framework including the background models and applied the analysis to actual flight data. Gerhard Heinzel and Min Zhong contributed to the interpretation of the results. Yihao Yan and Vitali M\"uller wrote the paper draft with input from all authors. All authors provided critical feedback and helped shape the research, analysis and manuscript. }
\\~\\
{\footnotesize
	\noindent
	\textbf{Data Availability Statement (DAS)}
	The level-1b data of the GRACE and GRACE Follow-On satellites analyzed in this article can be obtained from 
	NASA/PO.DAAC (https://podaac.jpl.nasa.gov/) or from ISDC (http://isdc.gfz-potsdam.de/grace-isdc/). }

\begin{table}[H]
	\centering	
	\caption{The mean value and peak amplitudes at once and twice  the orbital frequency ($f_\textrm{orb} =0.18$\,mHz) of the difference $\textrm{GRA\_KBR1B\_LTC}-c_0 \mathcal{T}_\textrm{DOWR}$, where $\mathcal{T}_\textrm{DOWR}$ is computed with different accuracy levels. See also fig.~\ref{fig:ComparisonGRACE}.}
	\begin{tabular}{llll}
		\hline\noalign{\smallskip}
		Constituents & mean &$f_\textrm{orb}$ & $2f_\textrm{orb}$ \\
		\noalign{\smallskip}\hline\noalign{\smallskip}
		$\mathcal T = \mathcal T_{SR}$&27\,nm &	0.2\,nm  & 1\,nm \\
		$\mathcal T = \mathcal T_{SR}+\mathcal T_{PM}$ &-331\,$\mu$m& 934\,nm & 90\,nm \\
		$\mathcal T = \mathcal T_{SR}+\mathcal T_{HM}$ &105\,nm& 0.5\,nm &  234\,nm \\
		$\mathcal T = \mathcal T_{SR}+\mathcal T_{SM}$ &27\,nm &	0.2\,nm  & 1\,nm \\
		$\mathcal T = \mathcal T_{SR}+\mathcal T_{PM}+\mathcal T_{HM}+\mathcal T_{SM}$ &-331\,$\mu$m&  934\,nm &325\,nm \\
		\noalign{\smallskip}\hline
	\end{tabular}
	
	\label{tab:2}       
\end{table}
\begin{table}[H]
	\centering	
	\caption{The mean value and peak amplitudes at once and twice  the orbital frequency ($f_\textrm{orb} =0.18$\,mHz) of the differences $\textrm{GFO\_KBR1B\_LTC} - c_0 \mathcal{T}_\textrm{DOWR}$ and $\textrm{GFO\_LRI1B\_LTC} - c_0 \mathcal{T}_\textrm{TWR}$ for different accuracy levels of $\mathcal{T}_\textrm{DOWR}$ and $\mathcal{T}_\textrm{TWR}$. See also fig.~\ref{fig:ComparisonGRACEFO}. }
	\begin{tabular}{lllllll}
		\hline \noalign{\smallskip} 
		\multirow{2}{*}{Constituents}                           & \multicolumn{3}{c|}{GFO/KBR}             & \multicolumn{3}{c}{GFO/LRI}              \\
		& mean  & $f_\textrm{orb}$ & $2f_\textrm{orb}$   & mean & $f_\textrm{orb}$ & $2 f_\textrm{orb}$  \\
		\hline\noalign{\smallskip}
		$\mathcal T = \mathcal T_{SR}$ &246\,$\mu$m & 1.3\,$\mu$m &56\,nm&246\,$\mu$m& 1.3\,$\mu$m &56\,nm       \\ 
		$\mathcal T = \mathcal T_{SR}+\mathcal T_{PM}$  &-35\,pm&3.6\,pm&6.6\,pm&2.1\,pm&14\,pm&16\,pm  \\ 
		$\mathcal T = \mathcal T_{SR}+\mathcal T_{PM}+\mathcal T_{HM}$ &57\,nm& 331\,pm&172\,nm&57\,nm&550\,pm            &172\,nm      \\ 
		$\mathcal T = \mathcal T_{SR}+\mathcal T_{PM}+\mathcal T_{SM}$ &-35\,pm & 3.6\,pm& 6.6\,pm&2.1\,pm& 14\,pm            &16\,pm        \\ 
		$\mathcal T = \mathcal T_{SR}+\mathcal T_{PM}+\mathcal T_{HM} +\mathcal T_{SM}$  &57\,nm& 331\,pm&172\,nm&57\,nm    &550\,pm&172\,nm       \\ 
		\hline\noalign{\smallskip}
	\end{tabular}
	\label{tab:3}       
\end{table}
\begin{table}[H]
	\centering	
	\caption{The mean value and peak amplitudes at once and twice  the orbital frequency ($f_\textrm{orb} =0.18$\,mHz) \reviewermark{of different  constituents in the LTC ($c_0 \cdot \mathcal{T}$). The values were computed using GRACE-FO GNI1B orbit data from 2019-02-05.}}
	\begin{tabular}{llllllllll}
		\hline \noalign{\smallskip} 
		\multirow{2}{*}{\reviewermark{Constituents}}                           & \multicolumn{3}{c|}{one way ranging}    & \multicolumn{3}{c|}{dual-one way ranging}          & \multicolumn{3}{c}{two way ranging}             \\
		& mean &$f_\textrm{orb}$ &$2f_\textrm{orb}$&mean&$f_\textrm{orb}$&$2f_\textrm{orb}$& mean&$f_\textrm{orb}$&$2f_\textrm{orb}$ \\
		\hline\noalign{\smallskip}
		$\mathcal T_{SR}$     &4.8\,m & 26\,cm& 4.5\,cm &-172\,$\mu$m &209\,$\mu$m&62\,$\mu$m& -123\,$\mu$m  &   419\,$\mu$m &124\,$\mu$m      \\
		$\mathcal T_{PM}$   &-246\,$\mu$m & 1.3\,$\mu$m &55\,nm   &-246\,$\mu$m  & 1.3\,$\mu$m & 55\,nm    &-246\,$\mu$m &  799\,nm    & 50\,nm     \\
		$\mathcal T_{HM}$ &57\,nm    &  1.7\,nm    & 169\,nm    &57\,nm  & 323\,pm     & 171\,nm    &57\,nm   &  805\,pm    & 171\,nm    \\
		$\mathcal T_{SM}$  &2.4\,pm    &  19\,fm    &   85\,fm  &-83\,am    & 0.8\,am      &   5.2\,am    &-63\,am   &  0.4\,am     & 0.07\,am      \\
		\hline\noalign{\smallskip}
	\end{tabular}
	\label{tab:GReffects}       
\end{table}


\bibliography{library}

\bibliographystyle{unsrt}

\end{document}